\documentclass[twocolumn,trackchanges]{aastex631}

\usepackage{multirow}

\received{July 15, 2022}
\revised{November 07, 2022}
\accepted{November 24, 2022}

\submitjournal{ApJ}

\shorttitle{Sustained chromospheric \& TR heating over a sunspot light bridge}
\shortauthors{Louis, Mathew, Bayanna, Beck \& Choudhary}

\begin{document}

\title{Sustained heating of the chromosphere and transition region over a sunspot light bridge}

\correspondingauthor{Rohan Louis}
\email{rlouis@prl.res.in}

\author[0000-0001-5963-8293]{Rohan E. Louis}
\affiliation{Udaipur Solar Observatory, Physical Research Laboratory\\
Dewali Badi Road, Udaipur - 313001, Rajasthan, India}

\author[0000-0002-9370-2591]{Shibu K. Mathew}
\affiliation{Udaipur Solar Observatory, Physical Research Laboratory\\
Dewali Badi Road, Udaipur - 313001, Rajasthan, India}

\author[0000-0001-5802-7677]{A. Raja Bayanna}
\affiliation{Udaipur Solar Observatory, Physical Research Laboratory\\
Dewali Badi Road, Udaipur - 313001, Rajasthan, India}

\author[0000-0001-7706-4158]{Christian Beck}
\affiliation{National Solar Observatory (NSO), \\
3665 Discovery Drive, Boulder, CO 80303, USA}

\author[0000-0002-9308-3639]{Debi P. Choudhary}
\affiliation{Department of Physics and Astronomy, \\
California State University, Northridge (CSUN), CA 91330-8268, USA}

\begin{abstract}
Sunspot light bridges (LBs) exhibit a wide range of short-lived phenomena in the chromosphere and transition region. 
In contrast, we use here data from the Multi-Application Solar Telescope (MAST), the Interface Region Imaging 
Spectrograph (IRIS), Hinode, the Atmospheric Imaging Assembly (AIA), and the Helioseismic and Magnetic Imager (HMI) 
to analyze the sustained heating over days in an LB in a regular sunspot.
Chromospheric temperatures were retrieved from the the MAST \ion{Ca}{2} and IRIS \ion{Mg}{2} lines by nonlocal 
thermodynamic equilibrium inversions. Line widths, Doppler shifts, and intensities were derived from the IRIS lines 
using Gaussian fits. Coronal temperatures were estimated through the differential emission measure, while the 
coronal magnetic field was obtained from an extrapolation of the HMI vector field. 
At the photosphere, the LB exhibits a granular morphology with field strengths of about 400\,G and no significant 
electric currents. The sunspot does not fragment, and the LB remains stable for several days. The chromospheric 
temperature, IRIS line intensities and widths, and AIA 171\,\AA\, and 211\,\AA\, intensities are all enhanced in the LB with 
temperatures from 8000\,K to 2.5\,MK. Photospheric plasma motions remain small, while the chromosphere and transition 
region indicate predominantly red-shifts of 5–-20\,km\,s$^{-1}$ with occasional supersonic downflows exceeding 100\,km\,s$^{-1}$. 
The excess thermal energy over the LB is about $3.2\times10^{26}$\,erg  
and matches the radiative losses. It could be supplied by magnetic flux loss of the sunspot ($7.5\times10^{27}$\,erg), 
kinetic energy from the increase in the LB width ($4\times10^{28}$\,erg), or freefall of mass along 
the coronal loops ($6.3\times10^{26}$\,erg). 
\end{abstract}





\section{Introduction}
\label{intro}
Light bridges (LBs) are bright, extended structures seen in the umbral
core of sunspots and pores. They exhibit a variety of morphologies 
resembling umbral dots, penumbral filaments, or quiet Sun granules, 
depending on their evolutionary phase 
\citep{1979SoPh...61..297M,2007PASJ...59S.577K,2012ApJ...755...16L}.
Penumbral LBs consist of small-scale barbs close to the edges of the 
filament \citep{2008ApJ...672..684R,2008SoPh..252...43L}, while
granular LBs  exhibit a dark lane along the central axis 
\citep{1994ApJ...426..404S,2003ApJ...589L.117B,2004SoPh..221...65L}.

LBs are conceived to be manifestations of large-scale magneto-convective structures 
\citep{1997ApJ...490..458R,2004ApJ...604..906R}, while 
\cite{1979ApJ...234..333P} and \citet{1986ApJ...302..809C} claim LBs to be 
field-free intrusions of hot plasma into the gappy
umbral magnetic field. \citet{1995A&A...302..543R}, \citet{1991ApJ...373..683L}, and 
\citet{1997ApJ...484..900L} have shown that the magnetic field within LBs 
is typically weaker and more inclined in comparison to the adjacent umbra.
\citet{2006A&A...453.1079J} suggested that the intrusion of hot, 
weakly magnetized plasma would force the adjacent umbral magnetic field
to form a canopy over the LB that could be a source for electric currents. 
Such a stressed magnetic topology in
LBs has often been cited as the driver for a number of reconnection-associated phenomena
such as small-scale jets \citep{2014A&A...567A..96L,2018ApJ...854...92T},
surges \citep{1973SoPh...28...95R,
2001ApJ...555L..65A,2015ApJ...811..137T,2016A&A...590A..57R}, strong brightenings
and/or ejections 
\citep{2008SoPh..252...43L,2009ApJ...696L..66S,2009ApJ...704L..29L}, 
as well as flares \citep{2003ApJ...589L.117B,2021ApJ...907L...4L}. 
Enhanced chromospheric activity, which is primarily 
transient in nature, appears to be an important characteristic 
of LBs \citep{2016AN....337.1033L}. 

It has been shown that the fragmentation of sunspots often occurs 
along LBs \citep{1987SoPh..112...49G,2012ApJ...755...16L}. While LBs signify 
convective disruption within sunspots with close similarities to the quiet-Sun 
at the photosphere \citep{1989SoPh..124...37S,1994ApJ...426..404S,2010ApJ...718L..78R}, 
their properties in the chromosphere and transition region appear distinct with 
enhanced emission and broad line widths \citep{2018A&A...609A..73R}. It has also been 
shown that LBs anchored to the penumbra can suppress the formation of coronal loops 
\citep{2021MNRAS.506L..35M}, which suggest their possible association to the 
large-scale magnetic topology of the active region. Recently, \citet{2020ApJ...905..153L} reported the 
formation of an LB through the large-scale emergence of a nearly horizontal magnetic 
structure within a regular sunspot. This emergence, which lasted about 13\,hr, was 
accompanied by strong temperature enhancements in the lower chromosphere all along 
the LB, which were produced by electric currents through ohmic dissipation 
\citep{2021A&A...652L...4L}. 
It is, however, 
unknown if there are other mechanisms that can heat the upper atmosphere of an LB over several 
days, particularly if the underlying structure has evolved sufficiently enough to facilitate vigorous 
convection similar to the quiet Sun.
We address the above issue in this article by 
investigating the source of sustained heating in the chromosphere and the transition region
over a granular LB in a regular sunspot over a duration of more than 48\,hr. 
Section~\ref{obs} describes the observations used. The data analysis is explained in Section~\ref{inversion}. 
The results are presented in Section~\ref{results} and discussed in Section~\ref{discuss},
while Section~\ref{conclude} provides our conclusions.


\begin{table*}[!ht]
\caption{Summary of observations from different instruments.} \label{tab01}
\begin{center}
\hspace{-90pt}
\begin{tabular}{c|c|c|cc}
\hline\hline
\multirow{2}{*}{Characteristics} & \multirow{2}{*}{MAST} & \multirow{2}{*}{HINODE} & \multicolumn{2}{c}{IRIS} \\ \cline{4-5}   
                                    &       &       & Raster & SJ \\
\hline\hline
\multirow{4}{*}{Date \& Time [UT]}  & 2019 May 14 & 2019 May 14          & \multicolumn{2}{c}{2019 May 14} \\ 
                                    & 04:20:23    & 17:49:02--18:21:21   & \multicolumn{2}{c}{00:24:50--01:51:15} \\ 
                                    & --          & 2019 May 15          & \multicolumn{2}{c}{2019 May 15} \\ 
                                    & --          & 11:50:05--12:22:24   & \multicolumn{2}{c}{11:57:46--12:46:43} \\ \cline{1-5}
\multirow{3}{*}{Lines/Wavelength [nm]}          & \multirow{3}{*}{\ion{Ca}{2}/854.2}         & \multirow{3}{*}{\ion{Fe}{1}/630} & \ion{Mg}{2}/280 & 279.6  \\
											  &					 &					   &\ion{C}{2}/133  & 133.0 \\
											  &					 &					   &\ion{Si}{4}/140 & 140.0 \\		 \cline{1-5}	
FOV $x-y$[\arcsec]                              & 190--190          & 152--164             & 112--175 & 167--175 \\
Spatial Sampling $x/y$ [\arcsec\,px$^{-1}$]  & (0.11)$^2$        & 0.29/0.32            & 0.35/0.33  & (0.33)$^2$ \\
Spectral Sampling [pm\,px$^{-1}$]            & 2.5               & 2.15                 & 5.09/2.58  &   --      \\
No. of scans/images                             & 1                 & 1                    & 1          & 80        \\                                      
\hline \hline
\end{tabular}
\end{center}
\end{table*}

\section{Observations}
\label{obs}
We study the leading sunspot in NOAA active region (AR) 12741 on 2019 May 14 and 15 
when it was at a heliocentric angle of about 17$^\circ$. We combine observations 
from several sources, which are described below.

\subsection{MAST Data}
\label{mast}
We utilize imaging spectroscopic observations using the narrowband imager 
\citep{2009ASPC..405..461M,2014SoPh..289.4007R,2017SoPh..292..106M} on the 
50\,cm Multi-Application Solar Telescope \citep[MAST;][]{2017CSci..113..686V}. 
The narrowband imager comprises two lithium 
niobate-based Fabry-P\'erot (FP) etalons, which are tuned by a kilovolt power supply 
along with a 16 bit digital-to-analog converter. Both FPs are housed in  
temperature-controlled ovens that provide a thermal stability of $\pm0.2^{\circ}$C. 
The FPs have a diameter of 60\,mm, a thickness of 226\,$\mu$m and 
577\,$\mu$m, a reflectivity $>$93\%, and are polished to an accuracy of $\lambda/$100.
At present, this instrument is being used to observe the photospheric 
\ion{Fe}{1} line at 617.3\,nm as well as the \ion{Ca}{2} line at 854.2\,nm, although 
any other wavelength can be observed by using an appropriate prefilter. In order to 
scan the above lines simultaneously, a dichroic beam splitter is placed after the 
low-resolution etalon so that only one FP is used for the \ion{Ca}{2} line while both 
FPs are used for the \ion{Fe}{1} line. This arrangement provides a full width at 
half-maximum (FWHM) and a free spectral range of 170\,m\AA\, and 7\,\AA,  
which is sufficient for the broad \ion{Ca}{2} line. A prefilter with an FWHM of 3\,\AA\, 
is used to suppress the secondary transmission peaks of the FP. The 2k$\times$2k 
filtergrams have a spatial sampling of about 0\farcs11\,px$^{-1}$ across a 200\arcsec\, 
field of view (FOV).

On 2019 May 14, we made three spectral scans of AR 12741 in the \ion{Ca}{2} line at 
04:20:23\,UT, 06:02:06\,UT, and 10:37:11\,UT with 81 wavelength points covering about 
$\pm$1\,\AA\, around the line center. The wavelength step was about 25.4\,m\AA\, and 
at each step 20 images were acquired. The total scan lasted about 4\,min. 
We only analyzed the first scan as the seeing was variable during the second and 
third scans.  

The filtergrams were corrected for darks, flats, and the field-dependent blue shift 
caused by the collimated mounting of the FP etalons 
\citep[see Sect.~2.1 of][]{cavallini2006}, which was about 27.4\,m\AA\, from the center 
to the edge of the FOV. 
The instrumental profile along with the prefilter curve was then determined by 
convolving the reference spectrum from the 
Fourier Transform Spectrometer \citep[FTS;][]{1984sfat.book.....K} atlas and matching it
to the observed mean quiet Sun spectrum.  For this 
study we select the central 1720$\times$1720 pixel region for analysis. In 
addition to the narrowband filtergrams, G-band and H$\alpha$ filtergrams with a spatial 
sampling of 0\farcs2\,px$^{-1}$ were also acquired from time to time. The H$\alpha$ filtergrams 
were obtained using a 0.5\,\AA\, wide Halle filter. 

\subsection{IRIS Data} 
\label{iris}
Raster scans and slit jaw (SJ) images  from the Interface Region Imaging Spectrograph 
\citep[IRIS;][]{2014SoPh..289.2733D} were also used along with the MAST \ion{Ca}{2} 
narrowband filtergrams. We primarily use the raster scans taken in the \ion{Mg}{2} k \& h 
lines at 280\,nm, the \ion{C}{2} 1334\,\AA\, and 1335\,\AA\, lines, and the \ion{Si}{4} 1394\,\AA\, 
and 1403\,\AA\, lines. The \ion{Mg}{2}, \ion{C}{2}, and \ion{Si}{4} lines form at a temperature 
of 10,000\,K, 30,000\,K, and 65,000\,K, respectively \citep{2018ApJ...854...92T}. 
The \ion{Mg}{2} k \& h lines correspond to the near-ultra violet (NUV) region, while 
the \ion{C}{2}, and \ion{Si}{4} lines correspond to the far
ultra violet (FUV) region.
The details of the IRIS datasets are summarized in Table~\ref{tab01}. IRIS raster scans use 
a 0\farcs35 wide slit, a spatial sampling of 0\farcs33\,px$^{-1}$ along the slit, and a spectral sampling 
of 50.9\,m\AA\, and 25.8\,m\AA\, in the NUV and FUV, respectively. The SJ images have a spatial 
sampling of 0\farcs33\,px$^{-1}$.

\subsection{Hinode Data} 
\label{hinode}
The vector magnetic field of the AR was obtained from observations made by the 
spectropolarimeter \citep[SP;][]{2001ASPC..236...33L,2008SoPh..249..233I} of the Solar 
Optical Telescope \citep{2008SoPh..249..167T} on board Hinode \citep{2007SoPh..243....3K}. 
Using the fast mode with 4.8\,s at each slit position, the SP mapped the AR from 
17:49:02--18:21:21\,UT on May 14 and 11:50:05--12:22:24\,UT on May 15. 
The four Stokes parameters of the \ion{Fe}{1} lines at 630\,nm were recorded by the SP 
with a spectral sampling, step width, and spatial sampling along the slit 
of 21.5\,m\AA, 0\farcs29, and 0\farcs32\,px$^{-1}$, respectively.
The SP FOV was 152\arcsec$\times$164\arcsec. 
Routines of the SolarSoft package \citep{2013SoPh..283..601L}
were used to reduce the observations to yield Level-1 data. 
Level-2 data were used for this study that comprise two-dimensional (2D) maps of the
magnetic field strength, inclination, azimuth, and line-of-sight
(LOS) velocity. These products were obtained from an inversion of the Stokes profiles using 
the MERLIN\footnote{MERLIN inversion products are provided by the 
Community Spectro-polarimetric Analysis Center at the following 
link – http://www.csac.hao.ucar.edu/csac} \citep{2007MmSAI..78..148L} inversion code.
The 180$^{\circ}$ azimuth disambiguation was carried out using the 
AMBIG\footnote{Code available at www.cora.nwra.com/AMBIG} code \citep{2009ASPC..415..365L}
based on the Minimum Energy Algorithm of \citet{1994SoPh..155..235M}. 
Following the disambiguation, the inclination and azimuth were transformed to the local 
reference frame. 
Table~\ref{tab01} summarizes the parameters from the various instruments.

\subsection{Solar Dynamics Observatory Data}
\label{sdo}
The data from the Solar Dynamics Observatory \citep[SDO;][]{2012SoPh..275....3P} consist 
of images from the Atmospheric Imaging Assembly \citep[AIA;][]{2012SoPh..275...17L}. 
We chose the images 
at a reduced cadence of 10\,minutes in the 171\,\AA\, and 211\,\AA\, extreme ultra violet (EUV)
channels that have a maximum temperature response from the transition region and corona, 
respectively. In addition, we also utilize
observations of the vector magnetic field from SDO's Helioseismic and Magnetic 
Imager \citep[HMI;][]{2012SoPh..275..229S} with a cadence of 12\,minutes and a spatial sampling 
of about 0\farcs5\,px$^{-1}$.


\section{Data Analysis}
\label{inversion}
The strategies for inferring the chromospheric temperature and other diagnostics are summarized below.

\subsection{NICOLE Inversions}
\label{nicole_inv}
The \ion{Ca}{2} spectra from the MAST narrowband imager were inverted using the
nonlocal thermodynamic equilibrium (NLTE) Inversion COde based on the Lorien 
Engine \citep[NICOLE;][]{2015A&A...577A...7S}. 
NICOLE inversions were carried out with two cycles, with a 
maximum of 25 iterations per cycle and using the FALC model \citep{1993ApJ...406..319F} as the 
initial guess atmosphere. The physical parameters resulting from the first cycle were used 
as inputs for the second cycle. In the first cycle of the inversion, temperature and LOS velocity, 
were perturbed with two nodes each with height-independent micro- and macro-turbulence. In the second 
cycle, the number of nodes for temperature and LOS velocity were changed to eight and four, 
respectively, with two nodes for micro-turbulence. 

\subsection{IRIS$^{\textrm{\footnotesize{2}}}$ inversions}
\label{iris_inv}
To infer the temperature in the upper photosphere and chromosphere from the IRIS 
\ion{Mg}{2} k \& h lines, we used the IRIS Inversion based on Representative 
profiles Inverted by STiC \citep[IRIS$^{\textrm{\footnotesize{2}}}$;][]{2019ApJ...875L..18S}.
IRIS$^{\textrm{\footnotesize{2}}}$ recovers the thermodynamic and kinematic properties of the
solar chromosphere by comparing the observed spectra to a database of representative profiles, 
which are averages of profiles sharing the same shape as a function of the wavelength. 
The atmospheric parameters for these representative profiles have in 
turn been derived from the STiC code \citep{2019A&A...623A..74D}, which 
synthesizes spectral lines in NLTE along with partial 
redistribution of scattered photons. The IRIS$^{\textrm{\footnotesize{2}}}$ database incorporates all the 
observational variants such as location on the solar disk, exposure time, and spatial sampling. 
The routines for recovering the physical parameters from a given raster scan are made available 
through the IDL distribution of SolarSoft. 

\subsection{Parameter Maps from IRIS}
\label{iris_param}
In addition to the IRIS$^{\textrm{\footnotesize{2}}}$ inversions we carried out single and double Gaussian 
fits to the \ion{Mg}{2} k line, \ion{C}{2} line at 1334\,\AA\, and the \ion{Si}{4} line at 1394\,\AA. These
fits are used to derive the peak intensity, Doppler shift, and line width over the 2D FOV.
The rest wavelength was determined from the average line profile in the smaller umbral core where the 
peak intensity was less than three times the root-mean-square noise level.

\subsection{Magnetic Field Extrapolation}
\label{extrapol}
We used a non-force-free field (NFFF) extrapolation technique \citep{2010JASTP..72..219H}
to infer the magnetic connectivity in and around the sunspot. This method is well suited to the 
high plasma-$\beta$ photospheric boundary \citep{2001SoPh..203...71G} and has been
successfully used in recent studies \citep{2020ApJ...899L...4Y,2021A&A...652L...4L}.

\begin{figure*}[!ht]
\centerline{
\includegraphics[angle=90,width=\textwidth]{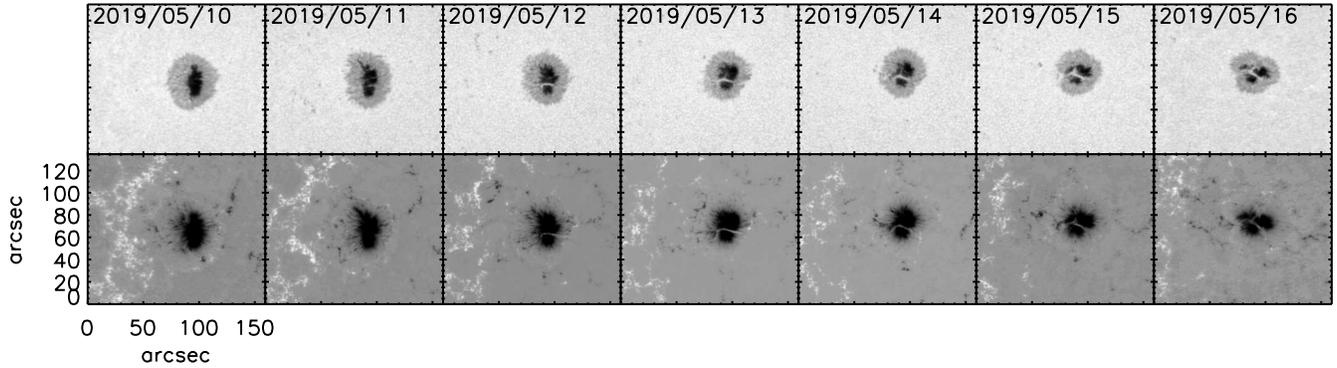}
}
\vspace{-230pt}
\caption{Evolution of NOAA AR 12741 as seen from HMI maps of the continuum intensity (top row) and the
vertical component of the magnetic field (bottom row). The maps correspond to 00:00\,UT.}
\label{fig01}
\end{figure*}

\subsection{Chromospheric Radiative Loss}
\label{radcalc}
The excess radiative loss in the LB over the quiet Sun (QS) was calculated using the procedure described in 
\citet[][their Sect.\,4.7]{2015A&A...582A.104R} for the MAST \ion{Ca}{2} line and the IRIS \ion{Mg}{2} k \& h lines.  
Following \citet{1984SoPh...90..205N}, the intensity of the QS at disk center at the respective central wavelengths 
was calculated and normalized by the factor arising from the heliocentric angle. The spectra in the QS were then 
integrated over a 1\,\AA\,band around the line core and subtracted from the same in the LB to yield the excess 
radiative loss.   

In order to calculate the total radiative loss in the chromosphere across the full spectrum, we scaled the values 
obtained from the \ion{Ca}{2} 854.2\,nm line to the remaining lines in the \ion{Ca}{2} IR triplet, the 
\ion{Ca}{2} K \& H lines as well as hydrogen Lyman $\alpha$, similar to the procedure described in \citet{2015A&A...582A.104R}. 
The scaling factors were chosen from \citet[][their Table 1]{2022A&A...665A..50Y} for a region over the polarity 
inversion line, which is quite similar to the conditions in the LB under study. We assume that the radiative loss 
in the \ion{Ca}{2} IR line at 854.2\,nm is a third of the loss of the whole  \ion{Ca}{2} IR triplet.

For a comparison to the observations we also synthesized\footnote{Calculation courtesy of J.~Jenkins/KUL.} the 
spectral lines of \ion{Ca}{2} IR, H \& K, \ion{Mg}{2} h \& k, and Ly$\alpha$ for a characteristic 
temperature stratification from a location inside the LB and the QS stratification with the Lightweaver 
code \citep{osborne+milic2021}. Radiative losses for the synthetic spectra were calculated by the same approach 
as the difference between LB and QS.


\begin{figure}
\centerline{
\includegraphics[angle=0,width=\columnwidth]{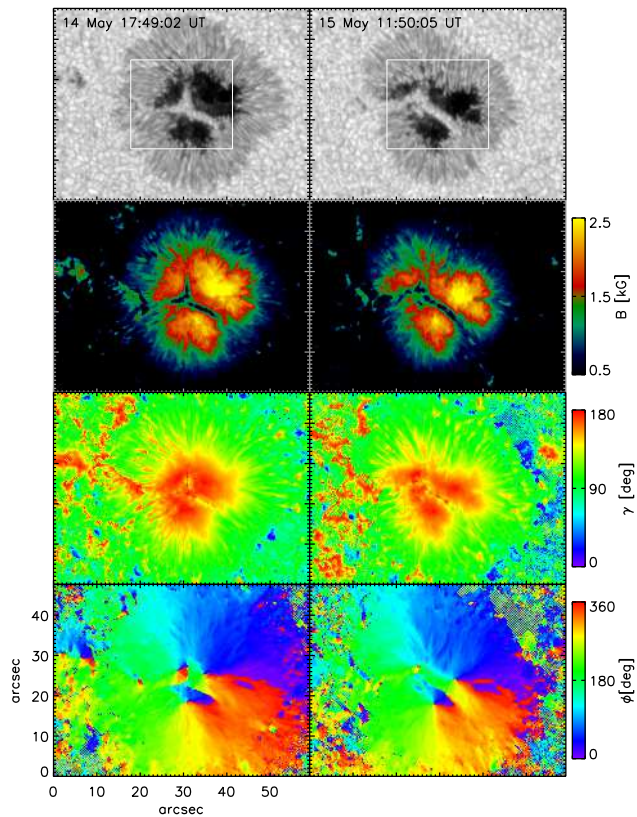}
}
\vspace{-25pt}
\caption{Hinode maps of the leading sunspot
in AR 12741 on 2019 May 14 (left) and May 15 (right). 
Top to bottom: continuum intensity $I_c$, field strength $B$, inclination $\gamma$, and azimuth $\phi$.
The white rectangle corresponds to the 
FOV shown in Figure~\ref{fig03}.}
\label{fig02}
\end{figure}

\section{Results}
\label{results}
\subsection{Evolution of Sunspot in AR 12741}
\label{morph}
Figure~\ref{fig01} shows the formation and evolution of the LB in the leading sunspot
in AR 12741. The leading sunspot appeared on the Eastern limb close to the end of May 6 
as a regular, unipolar spot without any conspicuous umbral intrusions. The formation of the LB began 
in the early part of May 11 with one arm extending from the western section of the 
umbra-penumbra boundary, eventually reaching the eastern umbral-penumbra boundary in about 
8\,hr. The onset of convection in the LB was seen in the latter part of May 13, with the subsequent
formation of a three-arm structure by May 15, which remained stable until the sunspot
traversed the western limb. During this period the sunspot did not fragment or decay into smaller
pores/spots and neither were there any significant changes in the global topology of the AR. 
The vertical component of the magnetic field in the figure shows that the LB stands
out in the umbral background with a relatively weaker amplitude. At HMI's spatial resolution we do not 
find any indication of opposite polarities in the LB during the lifetime of the sunspot.


\begin{figure}
\centerline{
\includegraphics[angle=0,width=\columnwidth]{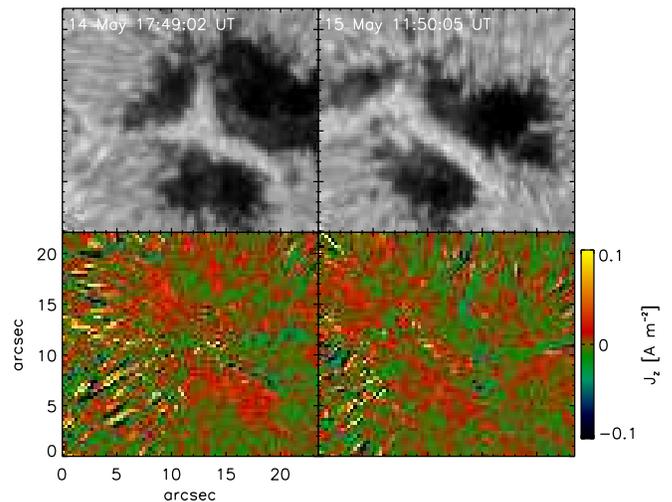}
}
\vspace{-145pt}
\caption{Magnified view of the LB in the SP data in the 
continuum intensity (top row) and the vertical component of electrical current 
density (bottom row) for the FOV marked by the white box in Figure~\ref{fig02}.}
\label{fig03}
\end{figure}


\begin{figure*}[!ht]
\centerline{
\hspace{270pt}
\includegraphics[angle=90,width=0.9\textwidth]{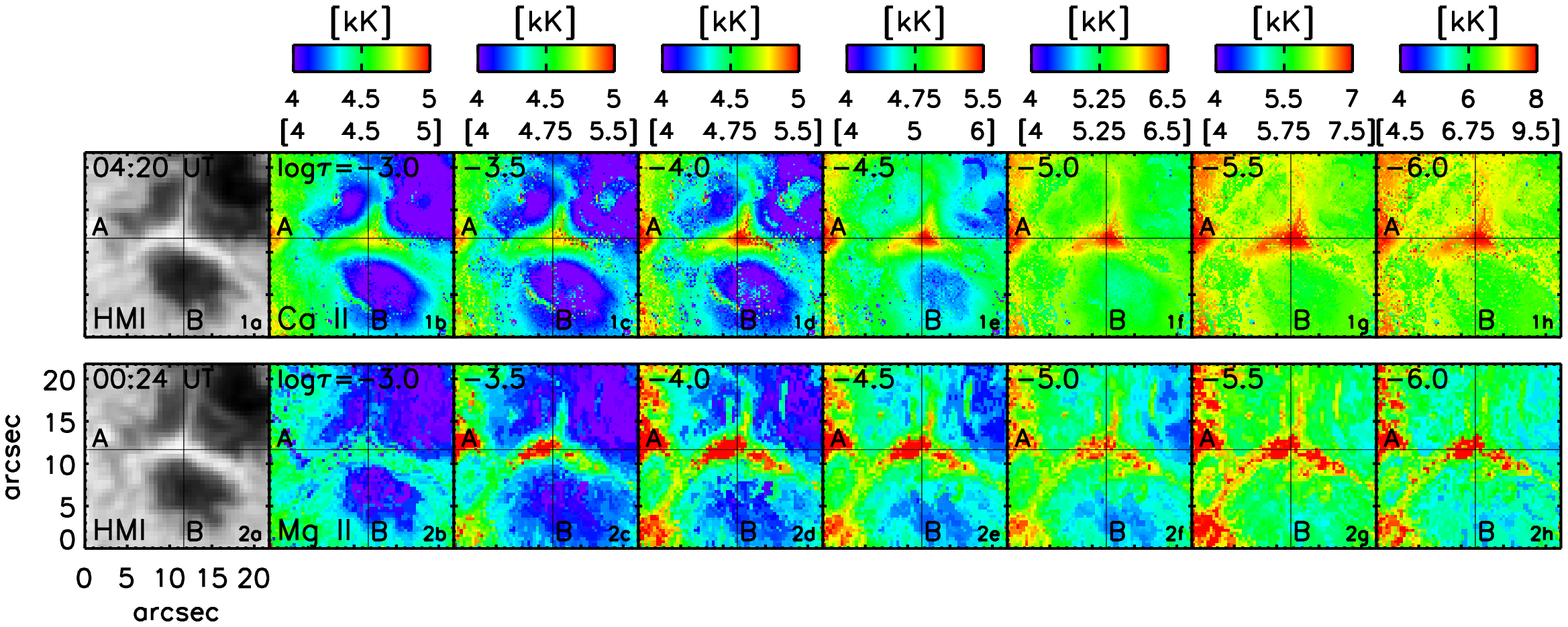}
\hspace{-70pt}
\includegraphics[angle=90,width=0.9\textwidth]{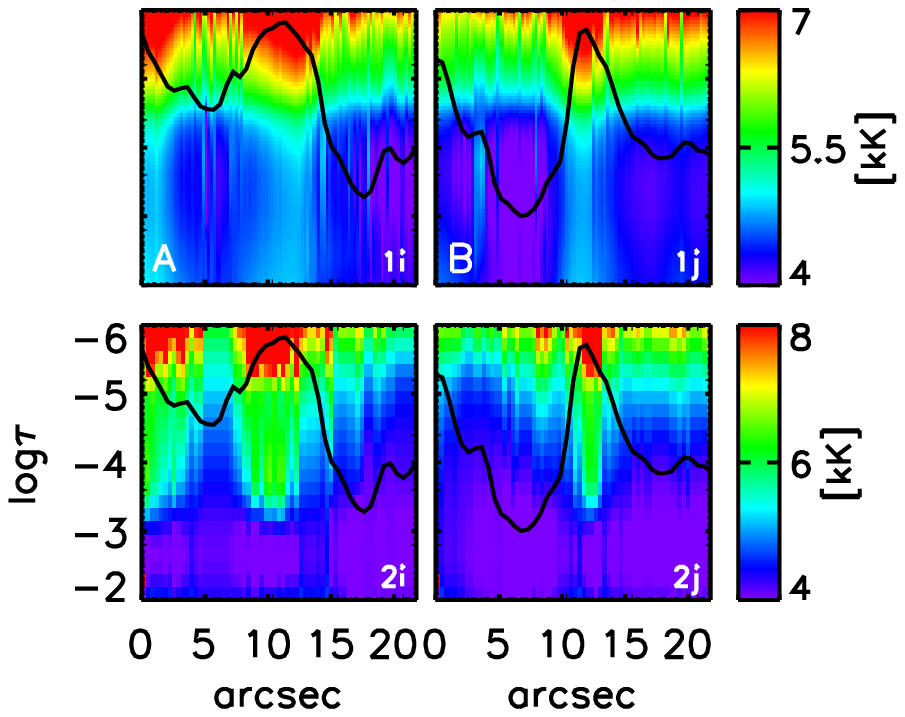}
}
\vspace{-175pt}
\caption{Comparison of temperature stratifications derived from the \ion{Ca}{2} and \ion{Mg}{2} lines
on May 14. Top row, from left to right: HMI continuum intensity (panel a), temperature derived from the MAST 
\ion{Ca}{2} line at different heights from $\log\tau=-3$ to $-6$ (panels b--h), temperature 
stratification along cut A and cut B in a 2D $x$--$\log\tau$ display (panels i and j). The black lines
in panels i and j correspond to the photospheric continuum intensity along cuts A and B, respectively.
Bottom row: same as above but for the temperature stratification derived from the IRIS
\ion{Mg}{2} lines. The temperature maps have been scaled to their respective color bars shown above the panels
where the values in parentheses correspond to the \ion{Mg}{2} lines.}
\label{fig04}
\end{figure*}

\subsection{Structure of Magnetic Field in LB}
\label{photos}
Figure~\ref{fig02} shows the Hinode continuum image of the sunspot wherein the
LB is seen to comprise bright, small-scale grains in all three arms with an absence of 
filamentary structures. At most locations, the continuum intensity in the LB is comparable 
to that in the quiet Sun.  The magnetic field is weakest along the axes of the LB, with 
minimum values ranging between 400\,G and 600\,G in the three arms. On the other hand, the 
field strength along the edges of the LB is about 1800--2000\,G, while in the umbra it 
ranges from about 2200\,G to 2700\,G. The magnetic field inclination increases as one moves 
from the edge to the axis of the LB with typical values in the interior ranging of about 
35$^\circ$--45$^\circ$ in all three arms.
As stated earlier, there are no indications of opposite polarities in the LB. The bottom row 
of Figure~\ref{fig02} reveals that for the most part the azimuth has a smooth radial arrangement 
in the sunspot, except in the proximity of the LB, which renders three azimuth centers in the 
respective umbral cores. As the sunspot is of negative polarity, the horizontal magnetic field 
is predominantly divergent along the LB axis and oriented toward the nearest umbral core.

Figure~\ref{fig03} shows a magnified view of the LB in the continuum intensity and the vertical 
component of the current density ($J_z$). The latter is extremely small in the LB, except at the 
locations along the axis of the LB, where the horizontal magnetic field appears to diverge 
into its respective umbral core. These locations are confined to individual pixels 
where $J_z$ can reach up to 0.15--0.25\,A\,m$^{-2}$, but these comprise only 3\% of the area of 
the LB. For the majority of the LB, however, the average values of $J_z$ are about 
0.02\,A\,m$^{-2}$ and do not stand out in the same manner as the currents in the penumbra 
that are relatively stronger by a factor of six. This obvious absence of strong electric currents 
in the LB is also seen in the Hinode map acquired on May 15. The absence of currents implies that 
heating by ohmic dissipation cannot be effective.


\begin{figure*}[!ht]
\centerline{
\includegraphics[angle=90,width=\textwidth]{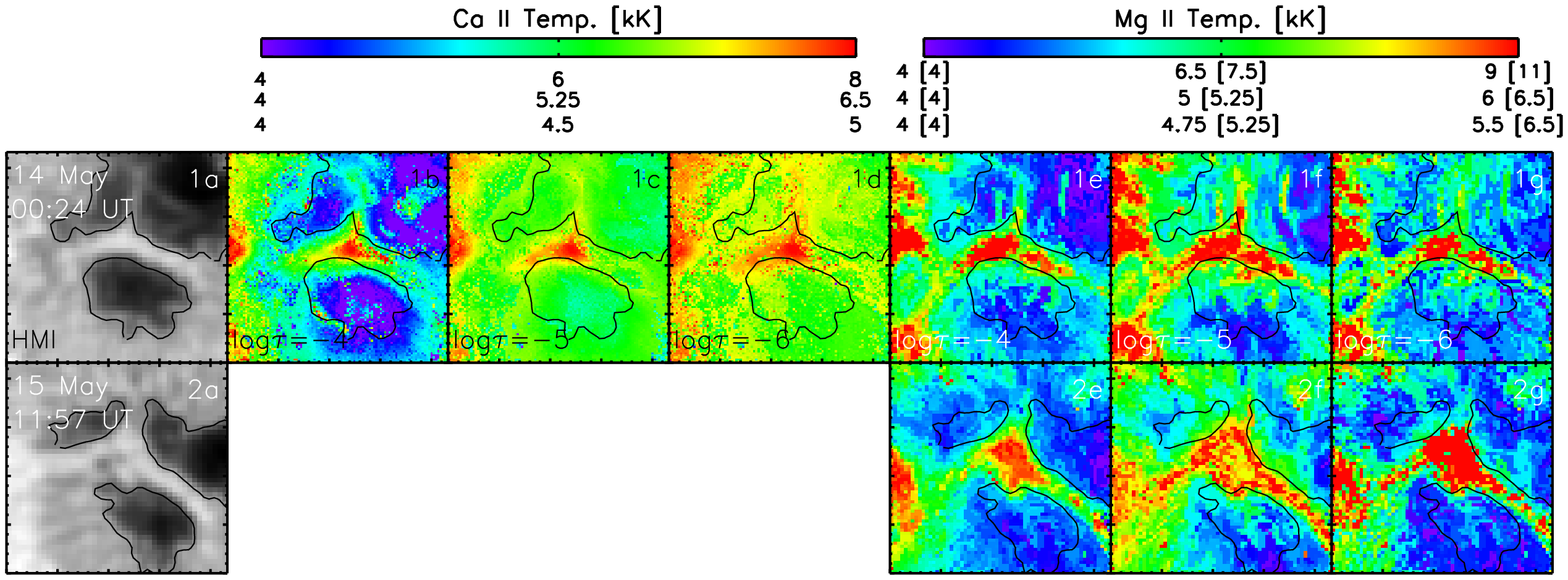}
}
\vspace{-245pt}
\centerline{
\includegraphics[angle=90,width=\textwidth]{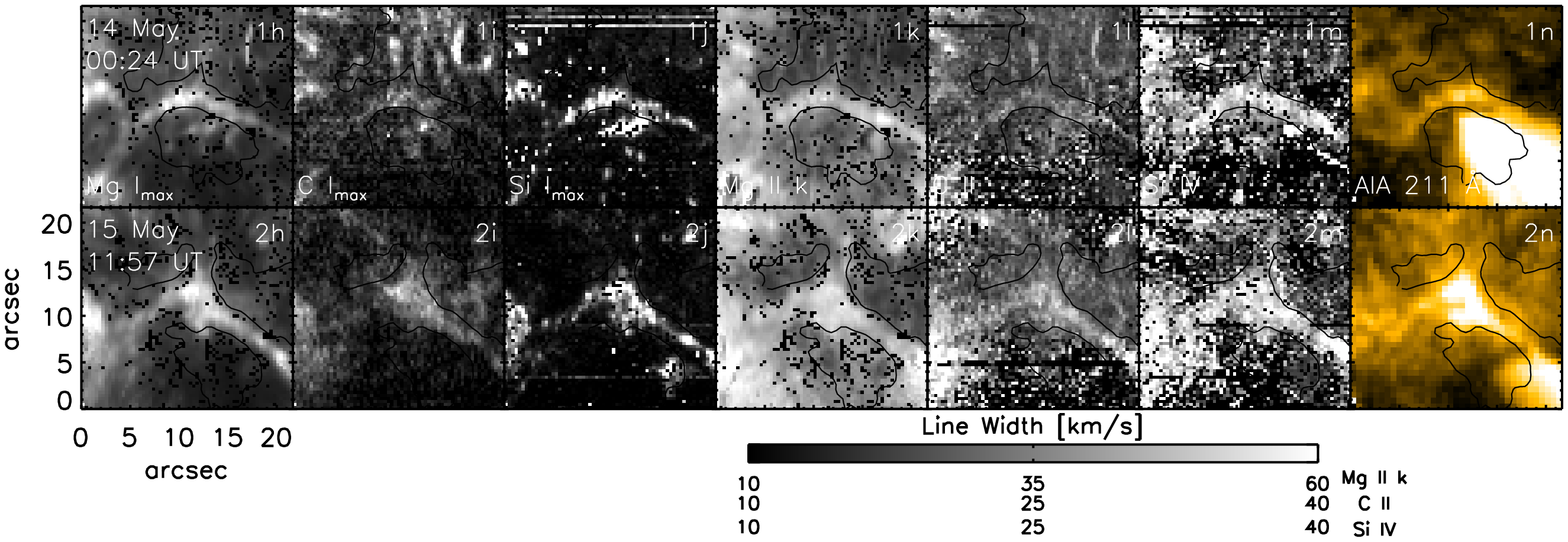}
}
\vspace{-145pt}
\caption{Maps of the temperature, peak intensity, and line width in the LB as a function of height. Top row, from left to right:
HMI continuum intensity (panel a), temperature derived from the MAST \ion{Ca}{2} line at $\log\tau=-4$, $-5$, $-6$ 
(panels b--d), and temperature derived from the IRIS \ion{Mg}{2} line at $\log\tau=-4$, $-5$, $-6$ (panels e--g). 
The temperature maps have been scaled to the corresponding color bars above 
the respective panels, with the numbers from the top to the bottom row below the color bar 
representing $\log\tau=-6$, $-5$, and $-4$, respectively. The temperature color bar for the IRIS \ion{Mg}{2} line 
is similar, with the numbers in the parentheses corresponding to the observations on 2019 May 15 at 11:57 UT 
for the maps in the second row.
Second row: the same as above on 2019 May 15 at 11:57 UT.
Third row: maximum line intensity from the IRIS \ion{Mg}{2} line, \ion{C}{2} line, \ion{Si}{4} line (panels h--j),
line width from the IRIS \ion{Mg}{2} line, \ion{C}{2} line, \ion{Si}{4} line (panels k--m), 
and AIA intensity at 171\,\AA (panel n) on 2019 May 14. Bottom row: the same as above on 2019 May 15 at 11:57 UT.
The black contours correspond to the HMI continuum intensity and outline the LB.}
\label{fig05}
\end{figure*}

\subsection{Enhanced, Persistent Brightness over LB}
\label{temp_lb}
Figure~\ref{fig04} shows the temperature maps of the LB as a function of height as derived from the spectral
inversions of the MAST \ion{Ca}{2} and IRIS \ion{Mg}{2} lines on May 14. The enhanced temperature in the LB 
appears over an extended height range of $-6.0<\log\tau<-3.5$ with the central junction of the LB being the 
hottest location. The average temperature at the central junction of the 
LB is 4960\,K, 6430\,K, and 7940\,K at $\log\tau=-4$, $-5$, and $-6$, respectively, as estimated from
the \ion{Ca}{2} line. These values in the LB exceed that of the umbra by 840\,K, 1165\,K, and 1315\,K at 
the above heights, respectively. The temperature maps from the \ion{Mg}{2} line exhibit similar values at $\log\tau=-4$
and $-5$ while at $\log\tau=-6$ the temperature is enhanced to 8300\,K. The 2D vertical cuts across the LB (panels i and j)
show that the thermal enhancement in the LB extends down to $\log\tau=-2$ in the \ion{Ca}{2} line while in the 
\ion{Mg}{2} line it is relatively higher at around $\log\tau=-3.5$. 
The temperature enhancement in the LB arises due to the reduced/suppressed absorption in the \ion{Ca}{2} line with 
the line core intensity being only 20\% smaller than the line wing intensity at about 1\,\AA\, (Figure~\ref{app_fig01} in 
the Appendix). On the other hand, the \ion{Mg}{2} k \& h lines comprise strong, compact emission features where the 
central reversals k3 and h3 are nearly as high as the k2 and h2 emissions. 

\begin{figure*}[!ht]
\centerline{
\hspace{40pt}
\includegraphics[angle=90,width=1.15\textwidth]{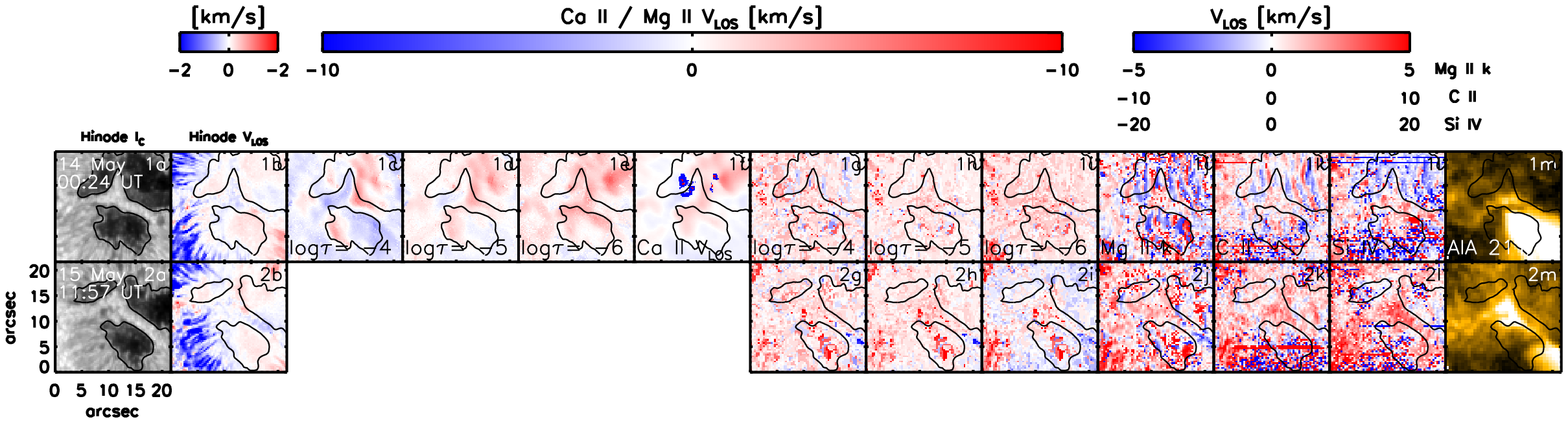}
}
\vspace{-280pt}
\caption{Maps of LOS velocity in the LB as a function of height. Top row, from left to right:
Hinode continuum intensity (panel a), Hinode LOS velocity (panel b), LOS velocity derived from the MAST 
\ion{Ca}{2} line at $\log\tau=-4$, $-5$, $-6$ (panels c--e), MAST \ion{Ca}{2} Dopplergram (panel f), 
LOS velocity derived from the IRIS \ion{Mg}{2} line at $\log\tau=-4$, $-5$, $-6$ (panels g--i), 
Doppler shift from the IRIS \ion{Mg}{2} line, \ion{C}{2} line, \ion{Si}{4} line (panels j--l)
and AIA intensity at 171\AA (panel m) on 2019 May 14. The velocity maps have been scaled to the corresponding 
color bars above the respective panels. The black contours, which outline the LB, correspond to the Hinode continuum 
intensity. Bottom row: the same on 2019 May 15.}
\label{fig06}
\end{figure*}

\begin{figure*}[!ht]
\vspace{-10pt}
\centerline{
\hspace{150pt}
\includegraphics[angle=90,width=1.3\textwidth]{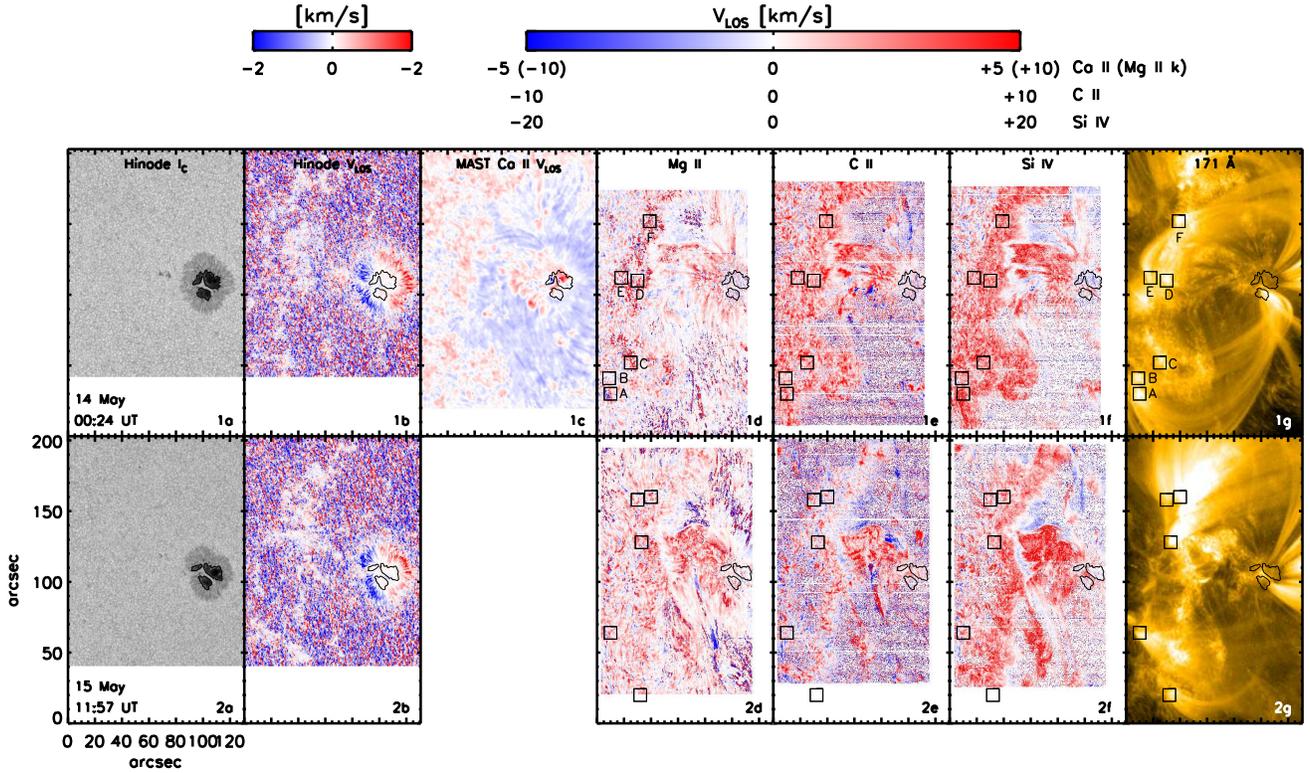}
}
\vspace{-175pt}
\caption{Maps of LOS velocity in AR 12741 as a function of height. From left to right:
Hinode continuum intensity (panel a), Hinode LOS Velocity (panel b), MAST \ion{Ca}{2} Dopplergram (panel c), 
Doppler shift from the IRIS \ion{Mg}{2} line, \ion{C}{2} line, \ion{Si}{4} line (panels d--f)
and AIA intensity at 171\AA (panel g). The velocity maps have been scaled to the corresponding color bars above 
the respective panels. The black contours, which outline the LB, correspond to the Hinode continuum intensity.
The black squares in panels 1g and 2g indicate the possible location of the outer footpoints of coronal loops 
that connect to the surroundings of the LB.}
\label{fig07}
\end{figure*}

\begin{figure*}[!ht]
\centerline{
\hspace{10pt}
\includegraphics[angle=90,width=1.1\textwidth]{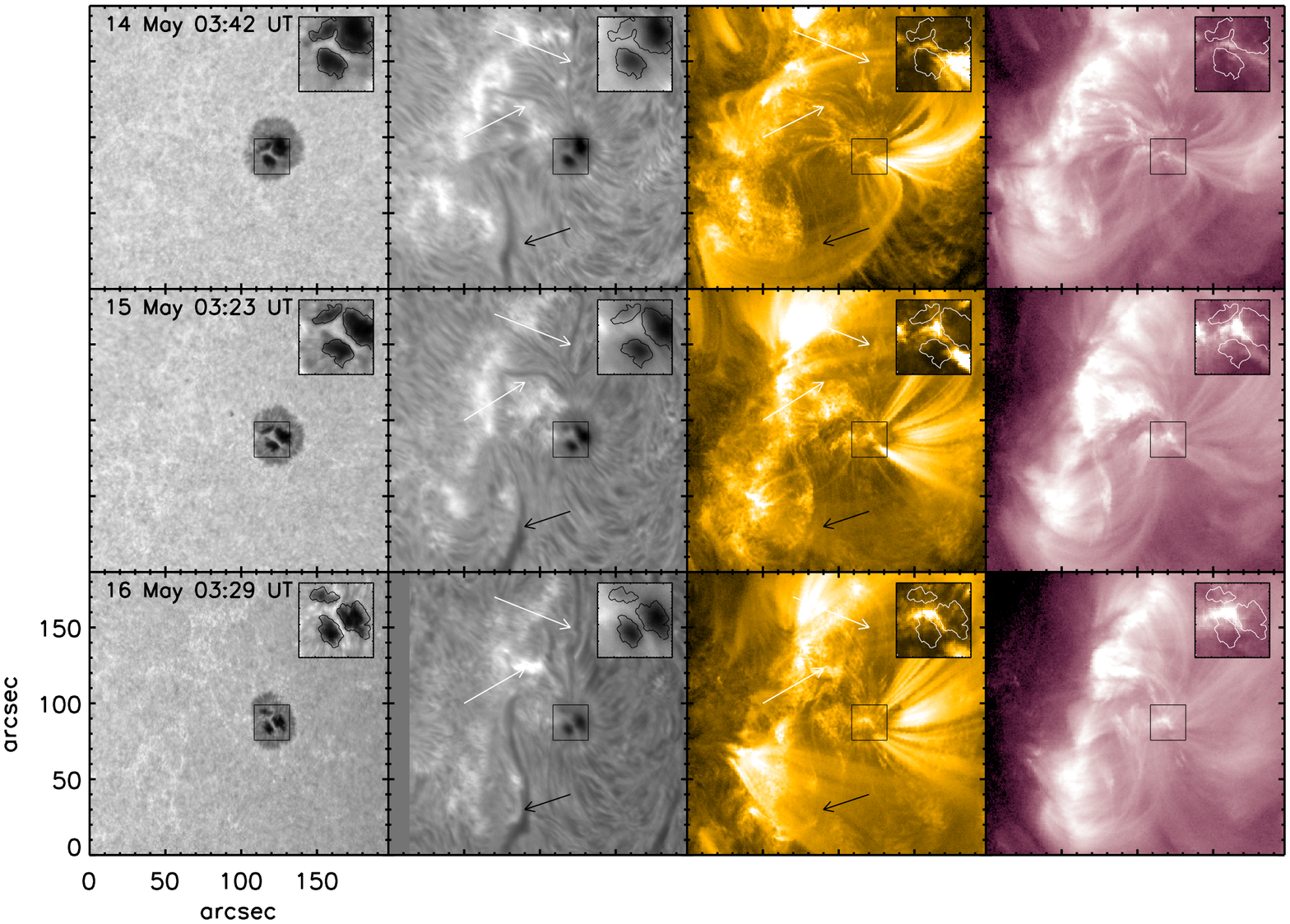}
}
\vspace{-60pt}
\caption{Photospheric, chromospheric, transition region, and coronal morphology in and around AR 12741 from 2019 May 14--16.
From left to right: G-band, H-$\alpha$ images from MAST, AIA images in the 171\,\AA, and 211\,\AA\, channels. The white and 
black arrows represent arch-filament systems and the AR filament, respectively.}
\label{fig08}
\end{figure*}

Figure~\ref{fig05} shows the temperature maps from the IRIS \ion{Mg}{2} line on May 14 and 15 along with the 
peak intensity and line width of the \ion{C}{2} and \ion{Si}{4} lines in the LB FOV. The temperature enhancement
over the LB persists over 36\,hr and coincides with the underlying photospheric morphology. A similar characteristic 
is seen in the peak intensity of the IRIS \ion{Si}{4} line while in the \ion{C}{2} line the LB is more diffused on 
May 14 than it is on May 15. The intensity in the LB is about 80\% of the emission in 
the opposite polarity network flux region as seen in the \ion{Mg}{2} and \ion{Si}{4} lines while in the \ion{C}{2}
it is about 24\% and 53\% on May 14 and May 15, respectively. 
The peak intensity of the \ion{Si}{4} line in particular also exhibits structures in the 
proximity of the LB that are associated with loops rooted at/close to the LB (Figure~\ref{app_fig02} in the Appendix).
The line widths estimated from the single Gaussian fit to the IRIS lines clearly 
trace the structure of the LB with values of 50\,km\,s$^{-1}$, 25\,km\,s$^{-1}$, and 30\,km\,s$^{-1}$ in the 
\ion{Mg}{2}, \ion{C}{2}, and \ion{Si}{4} lines, respectively, on May 14. These values nearly remain the same on 
May 15 for the \ion{Mg}{2} line while there is a marginal increase of about 5\,km\,s$^{-1}$ for the \ion{C}{2} 
and \ion{Si}{4} lines. The enhanced, persistent brightness in the LB can also be seen in the AIA 171\,\AA\, and 
211\,\AA\, images, which reflect conditions at transition region and coronal temperatures.
The enhanced intensity and line widths of the IRIS lines is also observed the following day on May 16, at 03:00\,UT.
The photospheric LB morphology can thus be traced up to the transition region and corona.


\subsection{Velocities with Height in LB}
\label{velo_lb}
Figure~\ref{fig06} shows the velocity in the LB in the photosphere, chromosphere, and transition region.
At the photosphere, the granular LB does not exhibit any strong red- or blueshifts with velocity values ranging 
between $\pm0.35$\,km\,s$^{-1}$. The velocities in the penumbra, however, are much stronger with the Evershed 
flow reaching 2\,km\,s$^{-1}$. The chromospheric velocities obtained from the inversion of the MAST \ion{Ca}{2} 
and IRIS \ion{Mg}{2} k \& h lines show that the LB is weakly redshifted by a few km\,s$^{-1}$ which 
nearly remains the same at heights of $\log\tau<-5$. The velocities obtained from the Gaussian fits to the IRIS 
lines (panels j--l) reveal that the LB is predominantly redshifted with values of 0.5\,km\,s$^{-1}$, 
2\,km\,s$^{-1}$, and 10\,km\,s$^{-1}$ in the \ion{Mg}{2}, \ion{C}{2}, and \ion{Si}{4} lines, respectively, on 
May 14. The \ion{Mg}{2} line was fitted with a double Gaussian while the \ion{C}{2}, and \ion{Si}{4} lines were 
fitted with a single Gaussian. While these redshifts in the IRIS lines persist in the LB on May 15, the values 
are increased to 1.5\,km\,s$^{-1}$, 5\,km\,s$^{-1}$, and 15\,km\,s$^{-1}$, respectively. The umbra in 
particular exhibits the saw-tooth pattern typically associated with shocks as seen in the \ion{Mg}{2} and 
\ion{C}{2} lines.

\begin{figure*}[!ht]
\centerline{
\includegraphics[angle=90,width=0.4\textwidth]{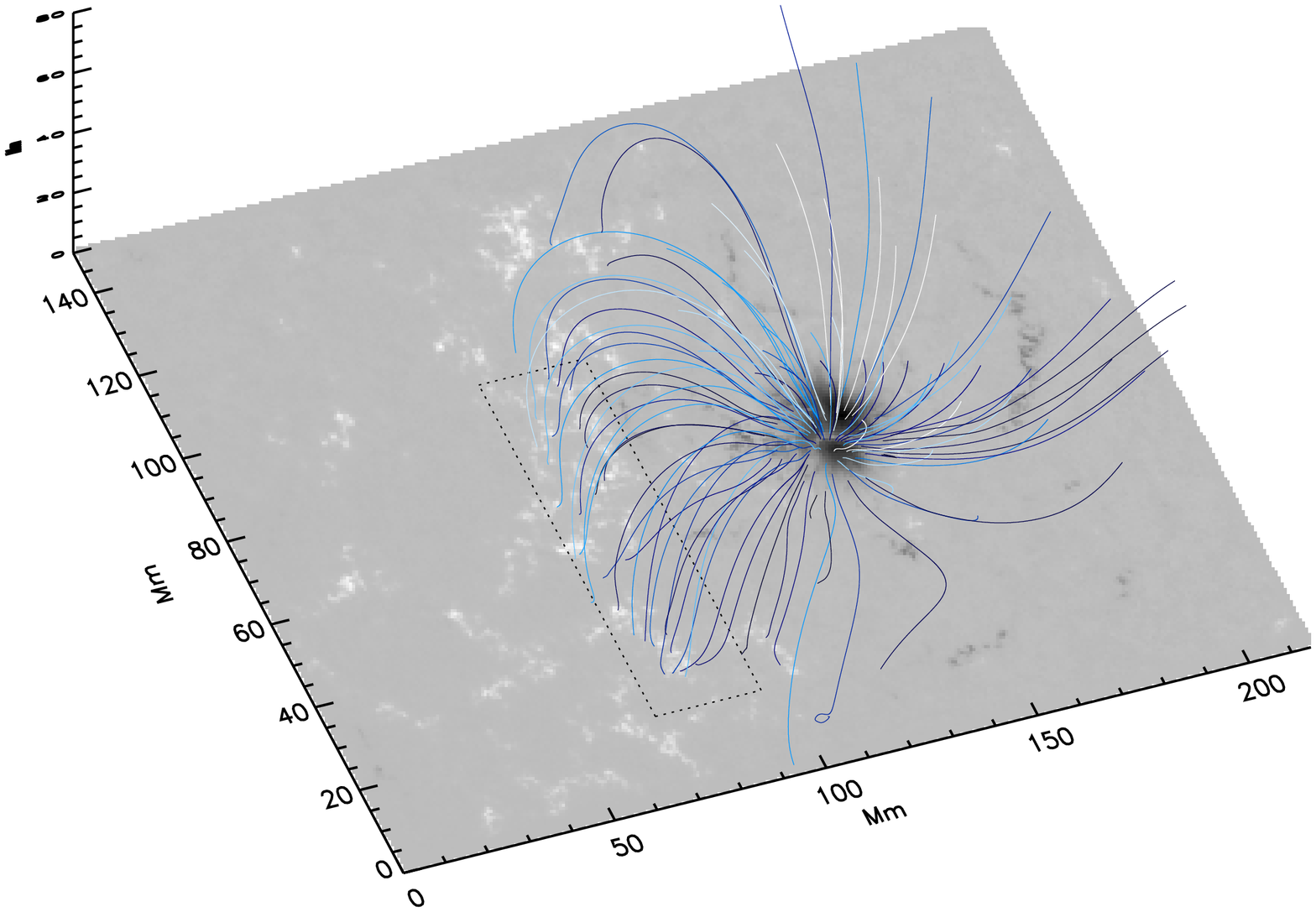}
\hspace{-50pt}
\includegraphics[angle=90,width=0.4\textwidth]{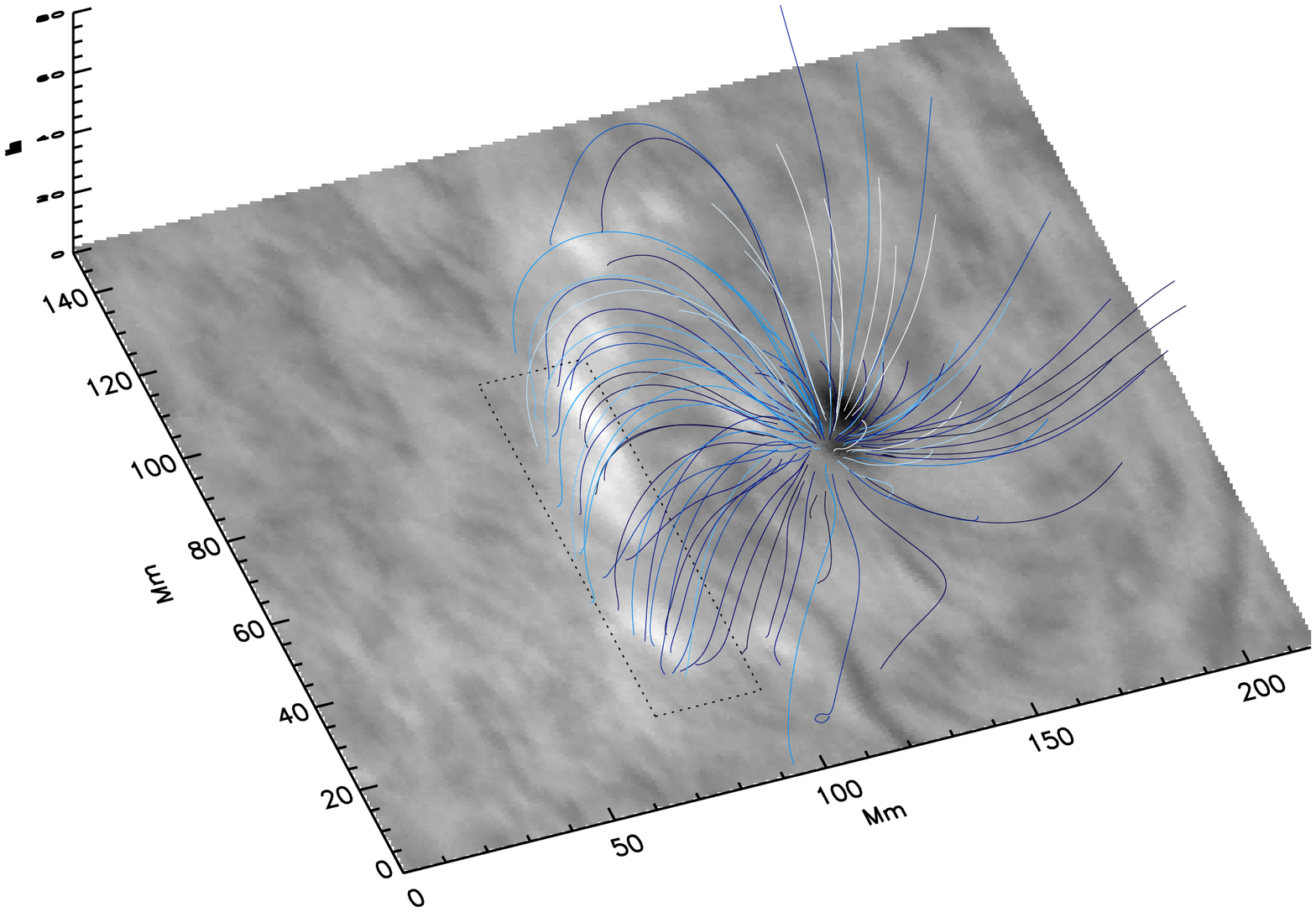}
\hspace{-50pt}
\includegraphics[angle=90,width=0.4\textwidth]{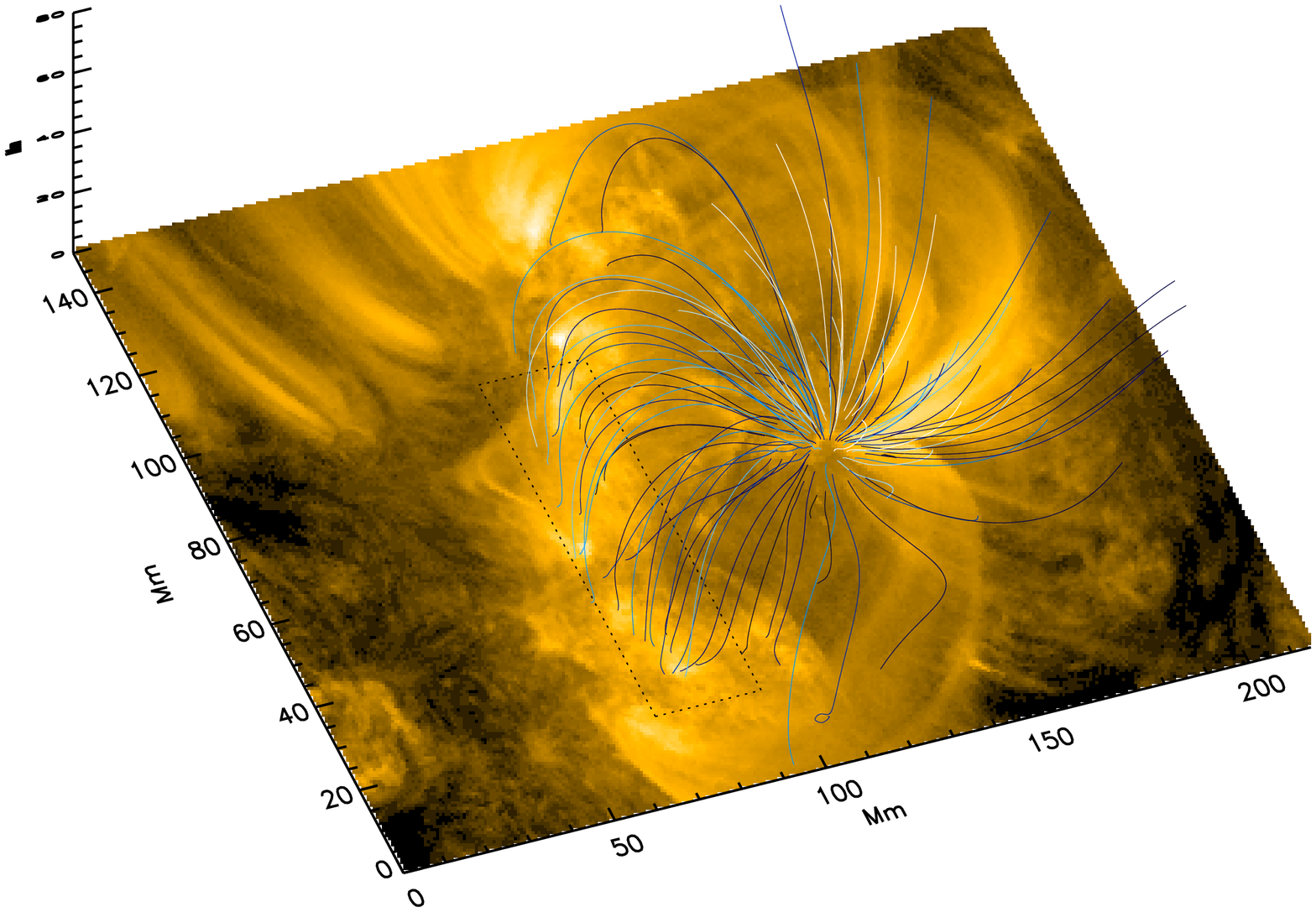}
}
\vspace{-50pt}
\centerline{
\includegraphics[angle=90,width=0.4\textwidth]{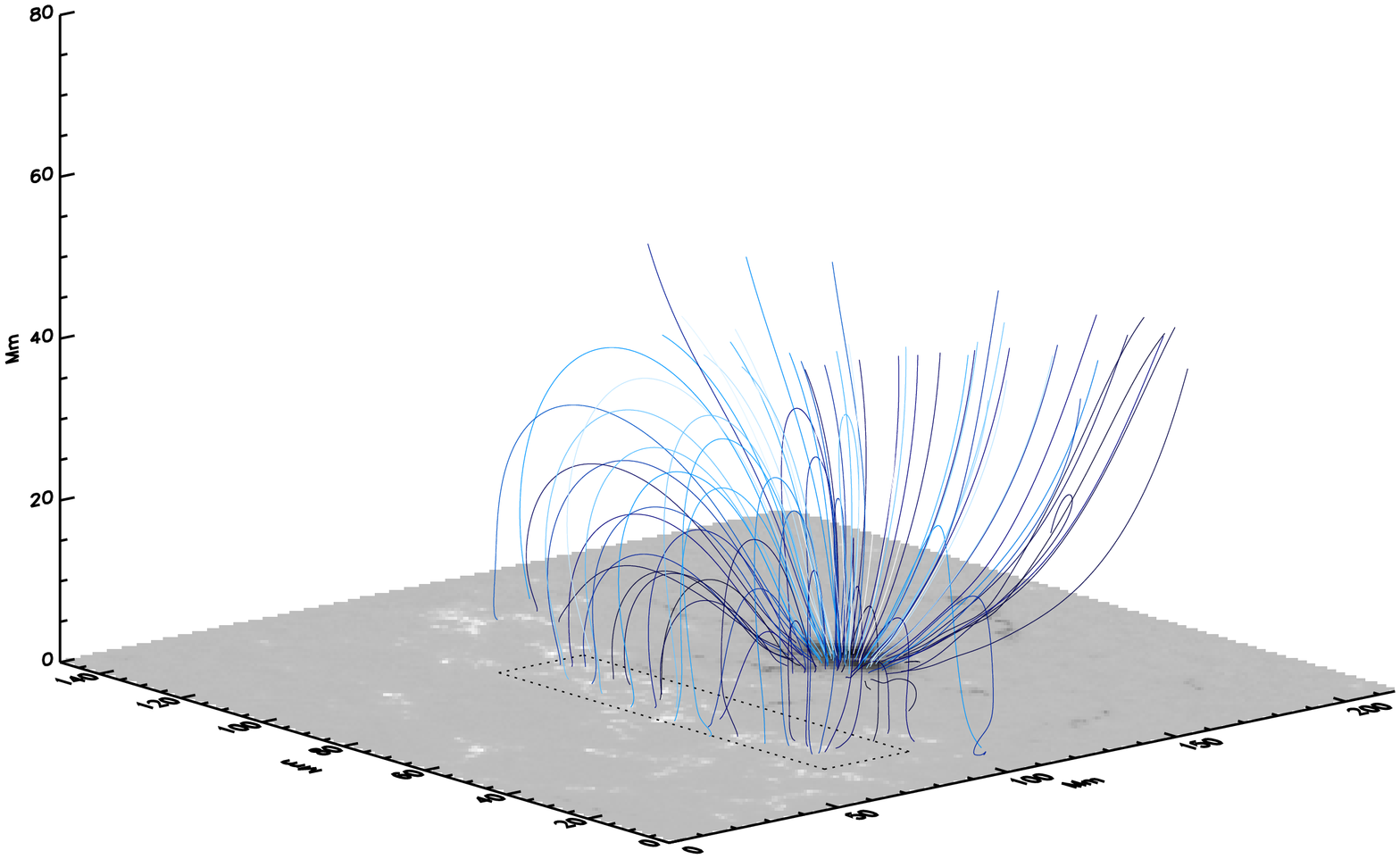}
\hspace{-50pt}
\includegraphics[angle=90,width=0.4\textwidth]{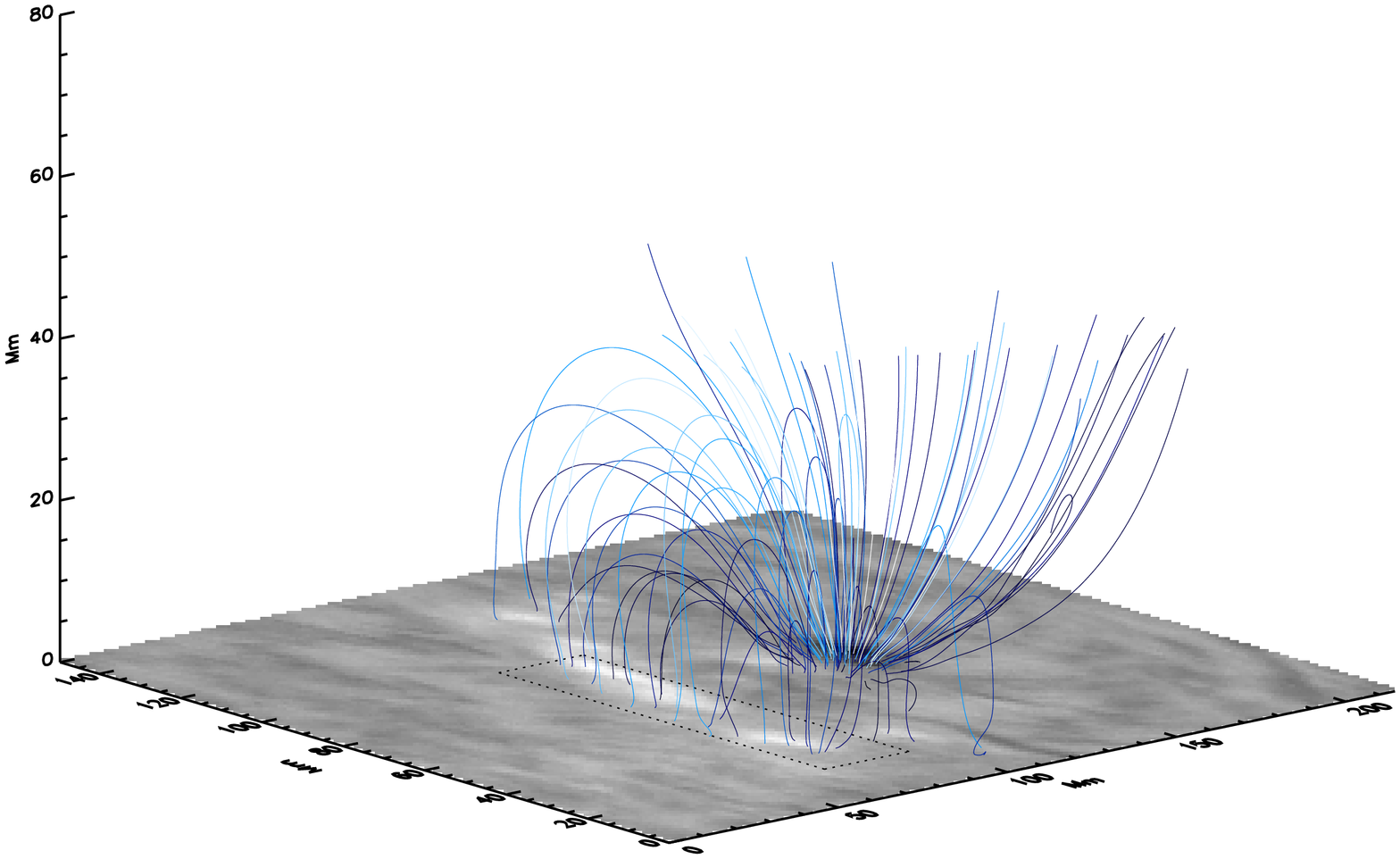}
\hspace{-50pt}
\includegraphics[angle=90,width=0.4\textwidth]{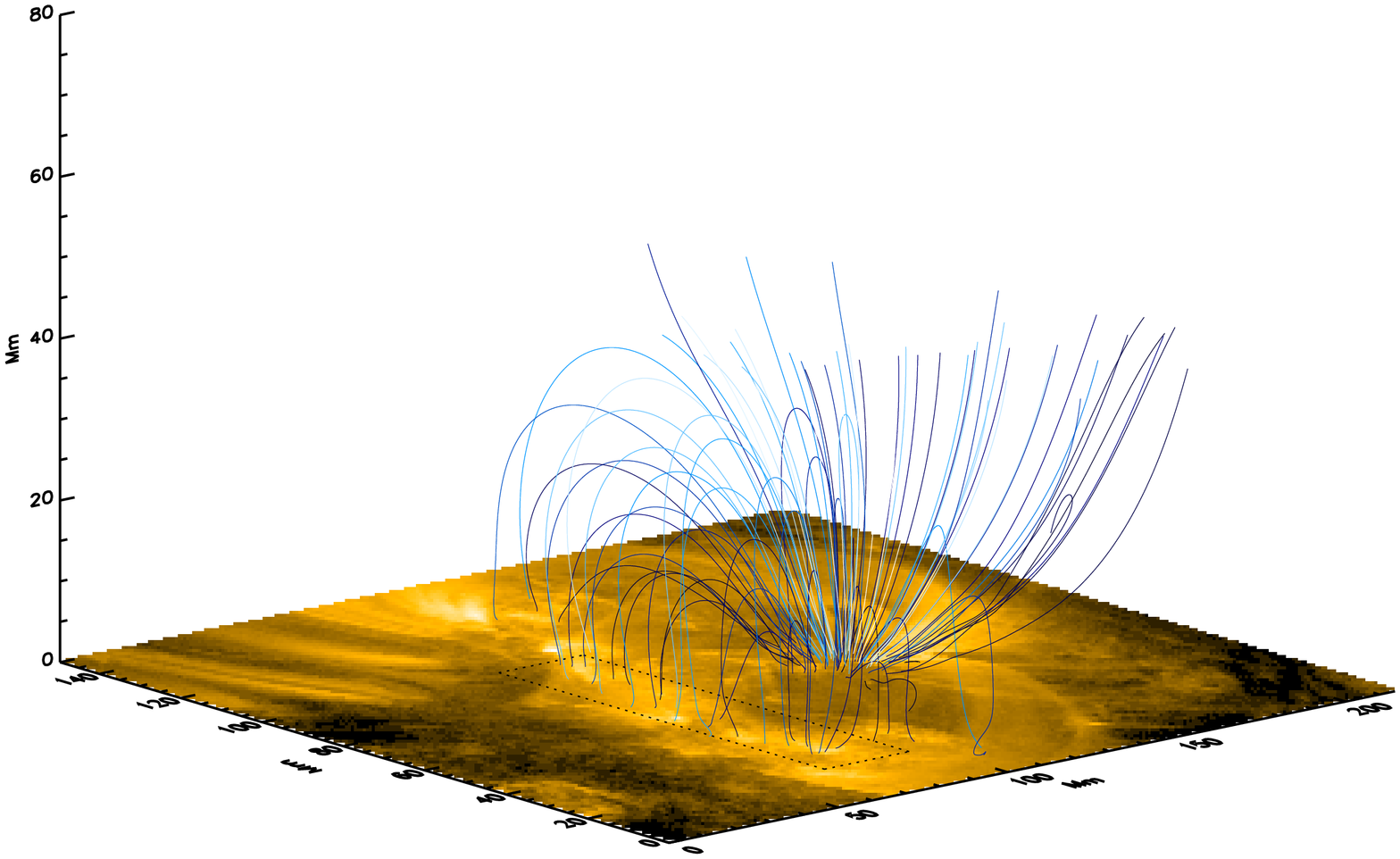}
}
\vspace{-10pt}
\caption{Global magnetic topology of NOAA AR 12741 from a NFFF extrapolation method
using the HMI vector magnetic field on May 14 at 04:24\,UT. 
The bottom boundary from left to right corresponds to the vertical component of the magnetic 
field from HMI, the GONG H-$\alpha$ filtergram, and an AIA 171\AA, image, respectively.
The black dotted rectangle encloses magnetic field lines from the sunspot
that connect to the network region of opposite polarity.}
\label{fig09}
\end{figure*}

\begin{figure}[!h]
\hspace{0pt}
\centerline{
\vspace{-50pt}
\includegraphics[angle=90,trim=80 50 100 50,width=1.3\columnwidth]{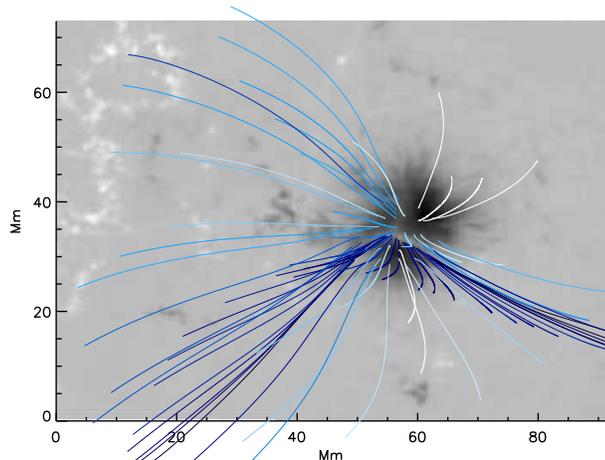}
}
\vspace{-20pt}
\caption{Magnified view of magnetic topology around the sunspot LB.}
\label{fig10}
\end{figure}

While the single Gaussian fits to the IRIS lines only show weak redshifts in the LB, an inspection of the spectra
emanating from the region in and around the LB reveal supersonic redshifts of about 150\,km\,s$^{-1}$ at the 
extended structure just south of the LB or next to it on May 14. These strong redshifts are associated with 
the large-scale loops ending in the sunspot (panels 7B--7E of Figure~\ref{app_fig02} in the Appendix). These 
spectra have been fitted with a double Gaussian to all the lines, which are also shown in Figure~\ref{app_fig02}. 
The high-speed downflows are primarily observed in the \ion{Si}{4} line and to very small extent in the \ion{C}{2} line 
(panel 4C). However, the downflows do not appear to persist in time and are greatly reduced as evident in the 
raster scans on May 14 at 13:21\,UT as well as on May 15 at 11:57\,UT. Figure~\ref{app_fig02} also shows the enhanced line 
width over the LB in all the IRIS lines, which remains a characteristic feature over the course of 36\,hr. 

Figure~\ref{fig07} shows the spatial distribution of velocities over the FOV of the AR. As stated earlier, 
the strongest velocities in the sunspot at the photosphere are associated with the Evershed flow, while in 
the chromosphere the inverse Evershed flow is observed in the superpenumbral region with values of about 
2\,km\,s$^{-1}$ and $-1.3$\,km\,s$^{-1}$ in the center-side and limb-side regions, respectively (panel 1c).  
In the IRIS \ion{C}{2} and \ion{Si}{4} the FOV is dominated by redshifts particularly in the network flux 
region and the arch-filament systems around the leading sunspot with values in the \ion{Si}{4} being the 
strongest and reaching about 20\,km\,s$^{-1}$. On the other hand, blueshifts appear sporadically in patches 
across the arch-filament systems as well as in the large filament east of the sunspot that extends southward, 
where the velocities are about $-3$\,km\,s$^{-1}$ and $-10$\,km\,s$^{-1}$, respectively, as estimated from the 
\ion{Mg}{2} line (panel 2d). 

We also visually identified the footpoints of the AIA 171\,\AA\ loops that begin from the LB and the sunspot and possibly 
end in the opposite polarity network flux region (squares in the panel). 

The majority of the footpoints are dominated by redshifts or extremely weak blueshifts apart from square A 
in panel 1d with a blueshift of about $-5$\,km\,s$^{-1}$. The transition region lines \ion{C}{2} and 
\ion{Si}{4} show dominantly redshifts all the time. 
The same trend is seen in the velocity maps on May 
15. However, unlike on May 14, the visible loops that begin at or close to LB do not appear to terminate at the 
opposite polarity network flux region. 

\begin{figure*}[!ht]
\centerline{
\hspace{10pt}
\includegraphics[angle=90,width=\textwidth]{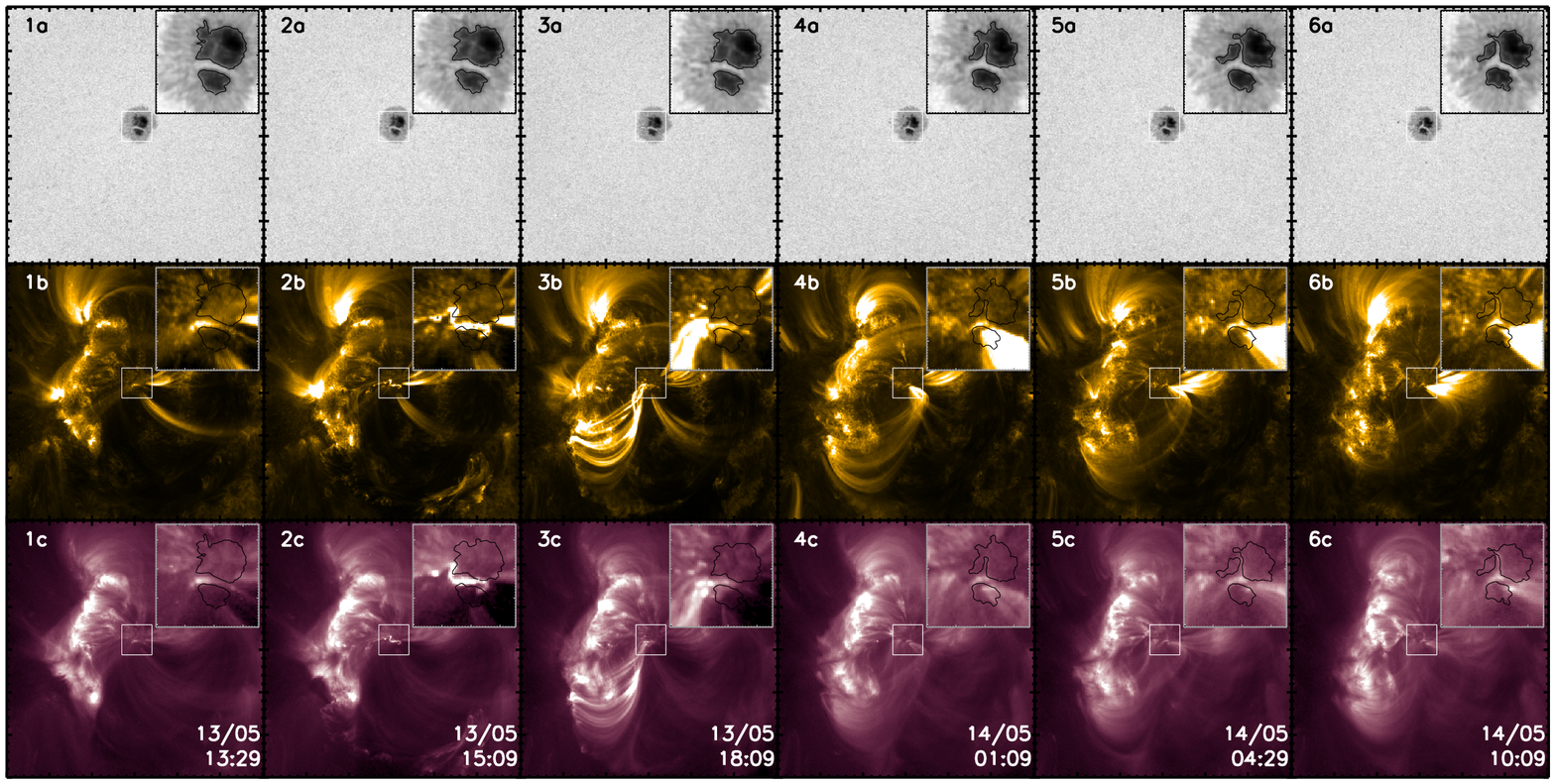}
}
\vspace{-140pt}
\centerline{
\hspace{10pt}
\includegraphics[angle=90,width=\textwidth]{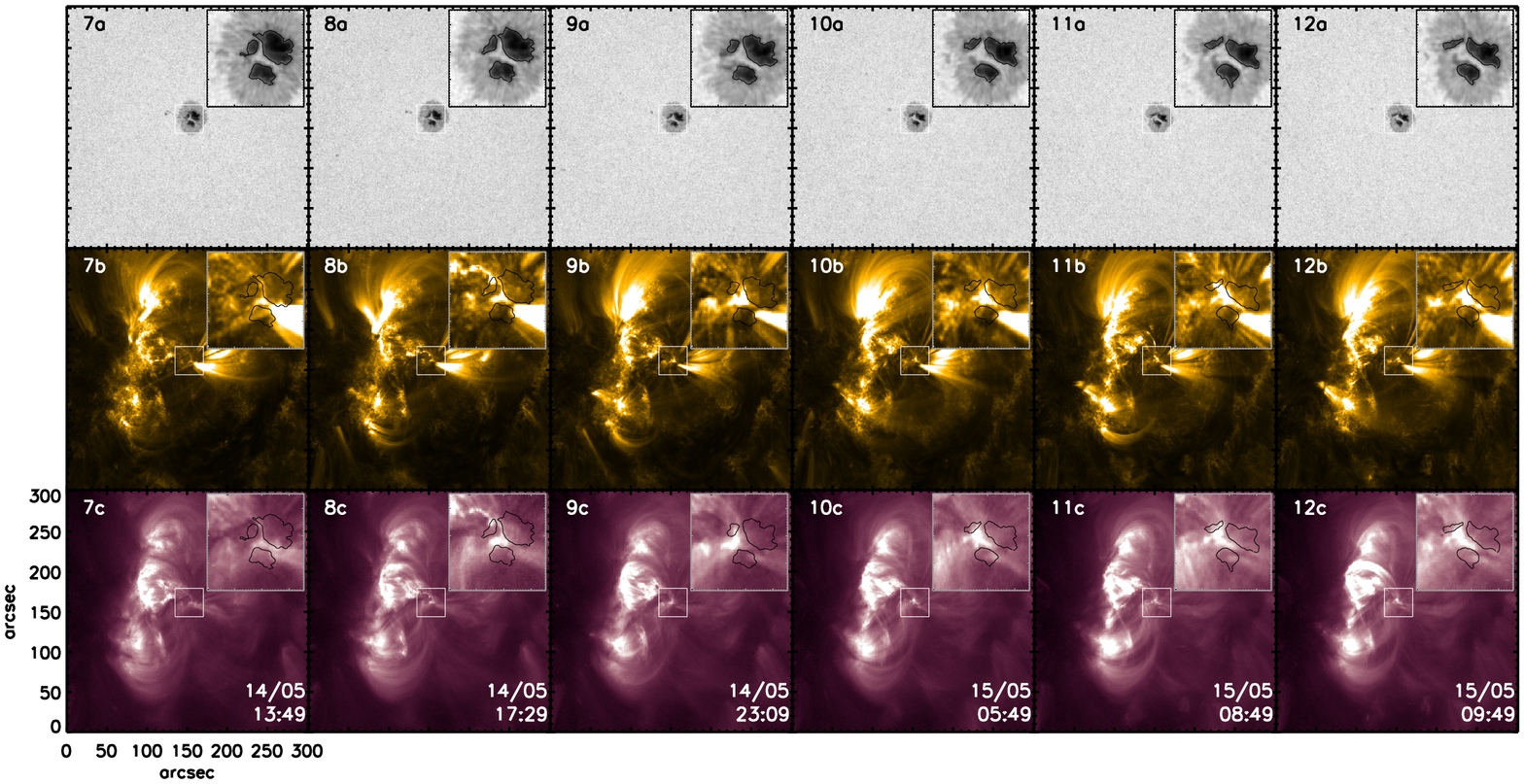}
}
\vspace{-120pt}
\caption{Evolution of the LB as seen in the transition region and corona. The top, middle, and bottom
panels correspond to the HMI continuum intensity and AIA 171\AA\, and 211\AA\,channels, respectively. 
The small white square in the middle of the FOV represents the LB whose magnified image is shown in 
the inset in the top right corner of the panel. The larger AIA images are clipped between 20 and 1500 
counts in both AIA channels. The insets on the other hand are 
scaled between 100--900 and 100--1000 counts in the 171 and 211\AA\,channels, respectively.}
\label{fig11}
\end{figure*}


\subsection{Global Topology and Morphology of AR}
\label{global_lb}
We now discuss the large-scale structures
in the chromosphere, transition region, and corona in the context of the temporal stability of the 
LB. Figure~\ref{fig08} shows that there are several, small 
arch-filament systems north of the leading sunspot that connect to the following polarity. In addition, a large 
filament located east of the sunspot extends southward beyond the MAST H$\alpha$ FOV and arches around to the 
west back toward the AR. These arch-filaments systems as well as the large AR filament also remain stable over 
a period of 36\,hr. The figure also shows that while the LB remains persistently bright in the AIA 171\,\AA\, 
and 211\,\AA\,images, there are large-scale loops that are always rooted at one end of the LB or close to it, 
which is seen from May 14 to May 16. In addition, these large-scale loops extend over and above the AR filament 
as observed on May 14.   

\begin{figure*}[!ht]
\centerline{
\includegraphics[angle=90,width=\textwidth]{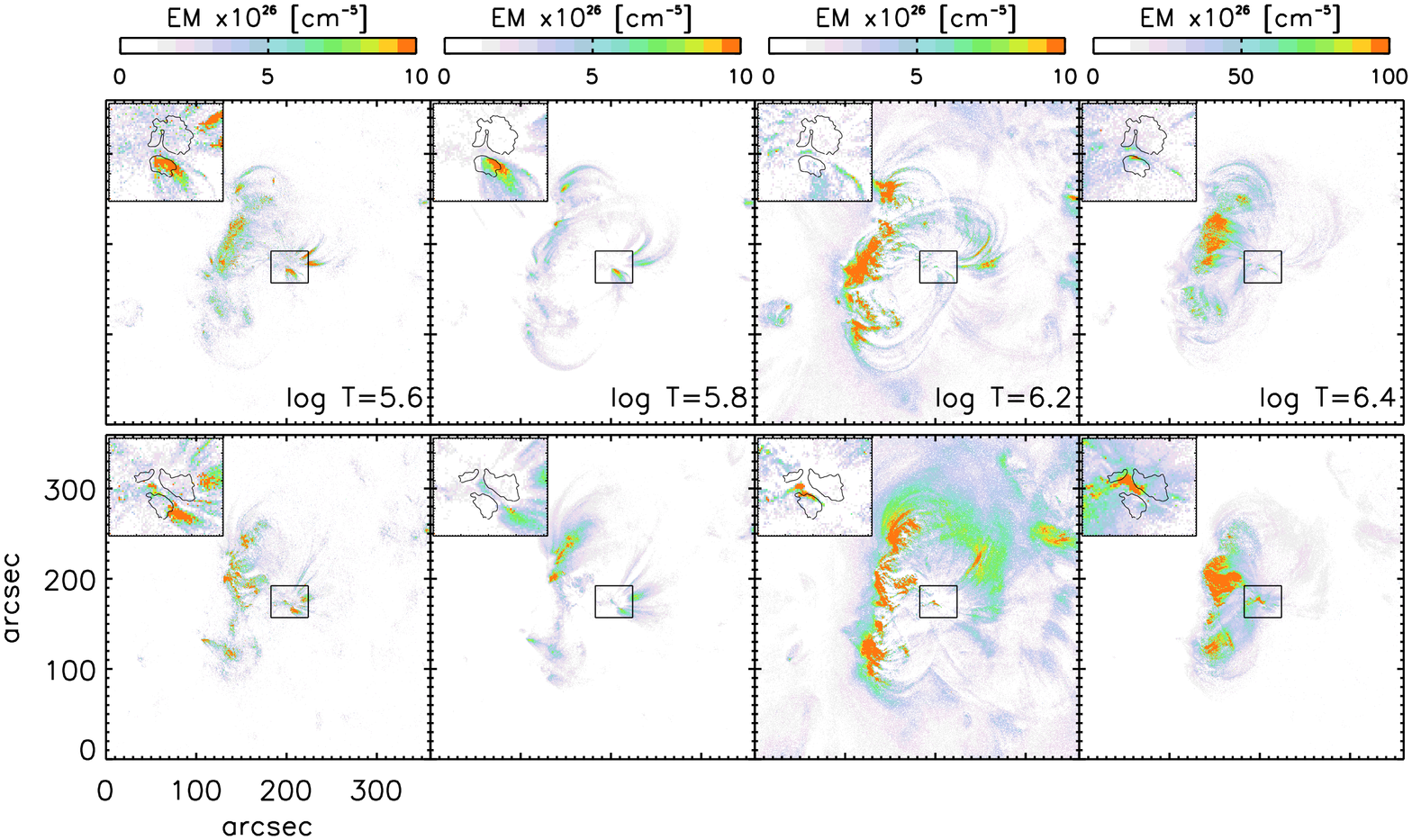}
}
\vspace{-70pt}
\caption{Emission Measure estimated from the AIA images using the 171, 211, 193, 335, 131, 94\,\AA\,channels 
at different temperatures. The top and bottom panels correspond to 2019 May 14 May at 00:24\,UT and 2019 May 15 
at 11:57\,UT. The magnified view of the light bridge is shown in the inset in the top left corner. The thin black 
contours correspond to the HMI continuum intensity.}
\label{fig12}
\end{figure*}

The global topology of the AR is further demonstrated from the extrapolation of the photospheric 
magnetic field using the NFFF technique as shown in Figure\ref{fig09} on May 14. The sunspot magnetic
field is consistent with a simple bipolar structure without any discernible signatures of twist. Field 
lines from the eastern penumbral region of the sunspot connect to the opposite polarity network flux region 
(black dotted rectangle in the figure) while those from the umbra and the western part of the sunspot 
comprise open field lines. The average height of the closed field lines connecting the sunspot to the 
following polarity is about 12.3\,Mm, while field lines starting from the inner penumbra can reach heights 
of up to 30\,Mm. An estimation of the loop height was independently made when the sunspot was very close 
to the western limb on May 19. An unsharp masked AIA 193\,\AA\, image provided a side view of the loops 
along the sky plane from which a height of about 13\,Mm was estimated (see Figure~\ref{app_fig03}
in the Appendix), which is in good agreement with those calculated from the extrapolated field lines. 
A zoom-in of the LB FOV (Figure~\ref{fig10}) shows the field being nearly vertical at the central part while toward 
the western end the field lines fan out with height. The LB thus seems to be related to features of the 
AR magnetic topology that govern its large-scale shape and evolution, at least in the sense of having their 
apparent footpoints in its vicinity.

Figure~\ref{fig11} shows the temporal evolution of the corona above the AR from May 13 to May 15. On May 13, 
a B3.5 class flare occurred at 15:02\,UT, with the peak at 15:52\,UT. The flare involved the eruption of 
the large AR filament south of the sunspot that was associated with a mass ejection that had a linear speed of 
312\,km\,s$^{-1}$, a position angle of 234$^\circ$
and was seen in the Large Angle Spectroscopic Coronagraph \citep[LASCO;][]{1995SoPh..162..357B} 
C2 FOV at 17:48\,UT as obtained from the Cactus \citep{2009ApJ...691.1222R} 
catalog\footnote{https://wwwbis.sidc.be/cactus/}. The erupting filament and the associated 
coronal dimming can be seen in the lower part of panel 2 of Figure~\ref{fig11}. The bright ribbon from the flare stretches 
all along the LB and stands out from the rest of the sunspot (panels 2b and 2c of Figure~\ref{fig11}). 
Similarly, the post flare loops from the ensuing eruption are rooted along the extended network flux region 
in the following group of sunspots of the AR while the other end is more confined and located at the eastern end of the LB 
(panel 3 of Figure~\ref{fig11}). There were no flares in the AR on May 14; however small-scale coronal activity was 
observed over the LB later in the day at around 17:30\,UT (panel 8). There were four flares
on May 15, which included a C2.0 class flare and three weak B-class flares. However, none of
these flares were eruptive and primarily involved the arch filament close to the northeastern  
periphery of the sunspot where one of the flares ribbons was seen. The other set of ribbons 
were located in the opposite polarity network flux region.

The AIA images clearly show the enhanced intensity over the LB as well as the presence of loops at or 
close to it, both of which persist over a duration of 3 days. 
Especially in panels 8--12 of Figure~\ref{fig11}, both AIA channels at 171\AA\, and 211\AA\, closely mimic 
the photospheric shape of the LB but at transition region heights.


\subsection{Energy Budget from Various Mechanisms}
\label{energy}
In this section we compare the energetics from various mechanisms/processes that could contribute to the  
sustained heating over the LB  for a duration of the observations.

\textit{(i) Thermal energy in the EUV: }The thermal energy 
emanating in the LB at EUV wavelengths can be estimated  as

\begin{equation} 
E_{th} = 3 n_e k_b T l^3,
\label{eq01}
\end{equation}

\noindent{}where $n_e$, $k_b$, $T$, and $l$ are the electron density, 
Boltzmann constant, temperature, and length scale over which the brightening in the LB is observed, 
respectively. To ascertain the temperature in the LB, we calculate the Differential 
Emission Measure \citep{2015ApJ...807..143C} from the various AIA channels, namely, 
171, 211, 193, 335, 131, 94\,\AA. Figure~\ref{fig12} shows the emission measure (EM)
at various temperatures, where the LB clearly stands out between $6.2\le \log{T} \le 6.4$.
The electron density $n_e$ can then be estimated as $n_e \approx \sqrt{EM/l}$. 
The value of the length scale $l$ is estimated from
the volume, using the area of the LB (25.8\,Mm$^2$ and 35.7\,Mm$^2$ on May 14 and 15, respectively) and a vertical
height of 4\,Mm. The electron density $n_e$ was computed using the mean DEM over the LB area and averaged over a 
temperature range of $\log T=6.2$ to $6.4$. With $T=2.5$\,MK, and $n_e = 1.8\times10^{9}$\,cm$^{-3}$ 
($2.5\times10^{9}$\,cm$^{-3}$), we obtain $E_{th}=1.9 \times 10^{26}$ ($3.7 \times 10^{26}$\,erg) in the LB on 
May 14 (15).

\begin{figure}[!h]
\centerline{
\includegraphics[angle=90,width=\columnwidth]{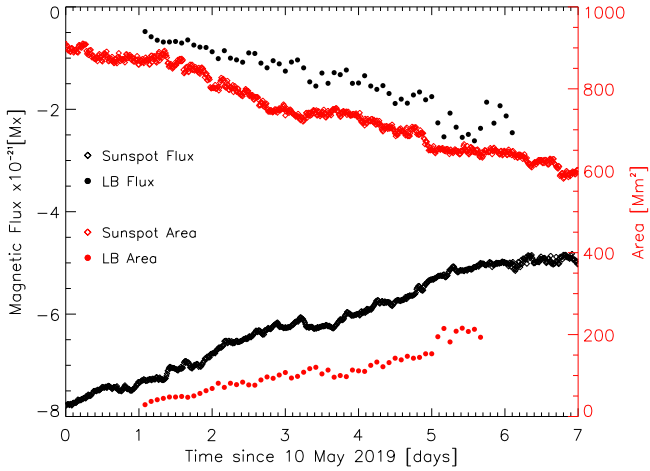}
}
\vspace{0pt}
\caption{Temporal evolution of the magnetic flux and area of the sunspot and light bridge. The y-axes on the left 
and right correspond to the flux and area, respectively. The light bridge flux and area have been enhanced by 
a factor 5 for better visibility.}
\label{fig13}
\end{figure}

\textit{(ii) Thermal Energy in the Visible and NUV: }The enhancement in internal energy can be computed from the temperature
stratification derived from the inversion of the MAST \ion{Ca}{2} and IRIS \ion{Mg}{2} lines following
\cite{2013A&A...553A..73B}.

\begin{equation}
\Delta E_{int} = \frac{R}{\mu(\gamma - 1)} \Delta A \sum_{i=z0}^{z1} \rho_i \Delta z_i \sum_{j,k} \left(T^{lb}_{i,j,k} - \overline{T}^{umb}_{i} \right),
\label{eq02}
\end{equation}

\noindent{}where $R=8.31$\,J\,mol$^{-1}$\,K$^{-1}$, $\mu=1.3$\,g\,mol$^{-1}$, $\gamma=5/3$, $\rho$ is the gas density, 
$\Delta A$ is the area of the pixel, 
$T^{lb}$ is the temperature in the LB, and $\overline{T}^{umb}$ is the average temperature in the umbra.
The summation with index $i$ is carried out from $z0$ to $z1$, which are the geometric heights at 
$\log\tau=-4$ and $\log\tau=-6$, respectively. The values of the geometrical height and gas density $\rho_i$ at 
different optical depth points were taken from the Harvard Smithsonian Reference Atmosphere 
\citep[HSRA;][]{1971SoPh...18..347G}. 
Indices $j$ and $k$ correspond to the spatial domain, while $\Delta z_i$ is the geometric height spacing 
between adjacent optical depth points. 
Equation~\ref{eq02} yields $4.7\times 10^{25}$\,erg and 
$8.6\times 10^{25}$\,erg for the thermal energy from the \ion{Ca}{2} and \ion{Mg}{2} lines, respectively.
$E_{th}$ is the enhancement in thermal energy over the surroundings of the LB, not the total 
energy. As it persists for days in a stationary way and the chromospheric relaxation time is on the order of 
a few minutes, the excess energy losses that lead to the enhancement must be replenished all the time by a 
continuing heating process.

\textit{(iii) Kinetic energy associated with LB expansion: }Figure~\ref{fig13} shows the 
temporal evolution of the sunspot flux and area, both of which decrease nearly linearly with time. The area 
of the LB on the other hand shows an increase with time, with a linear fit yielding a value of 
5.7\,Mm$^2$day$^{-1}$. Using the width of the LB of 3.3\,Mm on May 15, we can associate the linear expansion 
speed ($v_{frag}$) of the LB to the kinetic energy as 

\begin{equation}
E_{kin} = 0.5 A d \rho v_{frag}^2,
\label{eq03}
\end{equation}

\noindent{}where $\rho$ is the photospheric 
density, $A$ is the area of the LB, and $d$ is the depth to which the convective structure extends to, 
which we assume to be 6\,Mm \citep{2009ApJ...691..640R}. Using the above area increase and width of the LB, 
we obtain a value of 20\,m\,s$^{-1}$ for $v_{frag}$. We express the area and the density as a function of 
depth, namely, $\rho(z) = \rho_s\exp{(z/\tau_{\rho})}$ and $A(z) = A_s\exp{(-z/\tau_B)}$, where the suffix $s$ 
stands for the surface/photosphere and the scale heights in the two expressions correspond to the density and 
magnetic field. The values for $\tau_{\rho}$ and $\tau_B$ are 0.5\,Mm and 2\,Mm, respectively, while $\rho_s$ 
and $A_s$ are $10^{-7}$\,g\,cm$^{-3}$ and 35.7\,Mm$^2$, respectively. The kinetic energy can then be expressed 
as 

\begin{equation}
E_{kin} = \int_0^{d} 0.5 A(z) \rho(z) v_{frag}^2 dz,
\label{eq04}
\end{equation}

\noindent{}Using the above values, we obtain $E_{kin} = 3.9\times 10^{28}$\,erg.
For comparison, the convective energy of solar granulation with an  \textit{rms} velocity 
of 0.5\,km\,s$^{-1}$ \citep{2009A&A...507..453B,2013A&A...557A.109B} over the same area
as the LB and using Eqn.~\ref{eq04}, is about $2.4\times 10^{31}$\,erg, which is nearly 3 orders of magnitude larger
than the one from the expansion in the LB.

\begin{table}[!h]
\caption{Summary of energy estimates from different mechanisms for May 14 and May 15.} 
\label{tab02}
\begin{center}
\hspace{-30pt}
\begin{tabular}{c|cc}
\hline
\multirow{2}{*}{Mechanism} &   \multicolumn{2}{c}{Energy (erg)} \\ \cline{2-3}
& May 14 & May 15 \\
\hline\hline
Thermal Energy in EUV   & $1.9 \times 10^{26}$ & $3.7 \times 10^{26}$ \\
\hline
Thermal Energy in UV \& & $8.6\times 10^{25}$ & --\\
Visible                 & $4.7\times 10^{25}$ & --\\
\hline
Total Thermal Energy & $3.2\times 10^{26}$ & -- \\
\hline
Total Chromospheric & \multirow{2}{*}{$3 \times 10^{26}$}& \\
Radiative Loss\footnote{See Sect.~\ref{rad_loss}}      & &    \\
\hline\hline
Kinetic Energy          & $3.9\times 10^{28}$ & --\\
\hline
\multirow{2}{*}{Magnetic Flux}      & $7.5\times 10^{27}$ [S]& --\\
                        & $2.1\times 10^{26}$ [LB] &\\
\hline
Freefall               & $6.3\times10^{26}$ &  $8.7\times10^{26}$ \\
\hline
\end{tabular}
\end{center}
\end{table} 

\textit{(iv) Energy related to magnetic flux loss/gain: }As seen earlier, the sunspot loses magnetic flux 
at a rate of $\phi_t=1.8\times 10^{19}$\,Mx\,hr$^{-1}$, which was derived from a linear fit to the flux 
curve in Figure~\ref{fig13}. Similarly, the rate of flux increase in the LB is 
about $3.1\times 10^{18}$\,Mx\,hr$^{-1}$. The magnetic flux in the LB  increases because its area 
increases and the HMI data show a non-zero magnetic flux at those places.
The energy related to the loss of flux can be expressed as 

\begin{equation}
E_{flux} = \frac{1}{8\pi}\frac{(\phi_t t)^2}{h},
\label{eq05}
\end{equation}

\noindent{}where $t$ is the time scale over which the flux lost/gained can supply 
the energy ($\sim$10\,min) and $h$ is the chromospheric heating height scale  
\citep[$\sim$500\,km;][]{2018A&A...615L...9C}. The loss of flux in the sunspot provides an energy of $7.5\times 10^{27}$\,erg, 
while that gained in the LB is about $2.1\times 10^{26}$\,erg.
 
\textit{(v) Free fall energy: }The free fall energy of plasma draining down a loop 
from a height $h$ in the corona can be expressed as 

\begin{equation}
E_{fall} = \rho A g_\odot h^2,
\label{eq06}
\end{equation}

\begin{figure}[!h]
\centerline{
\hspace{0pt}
\includegraphics[angle=90,width=\columnwidth]{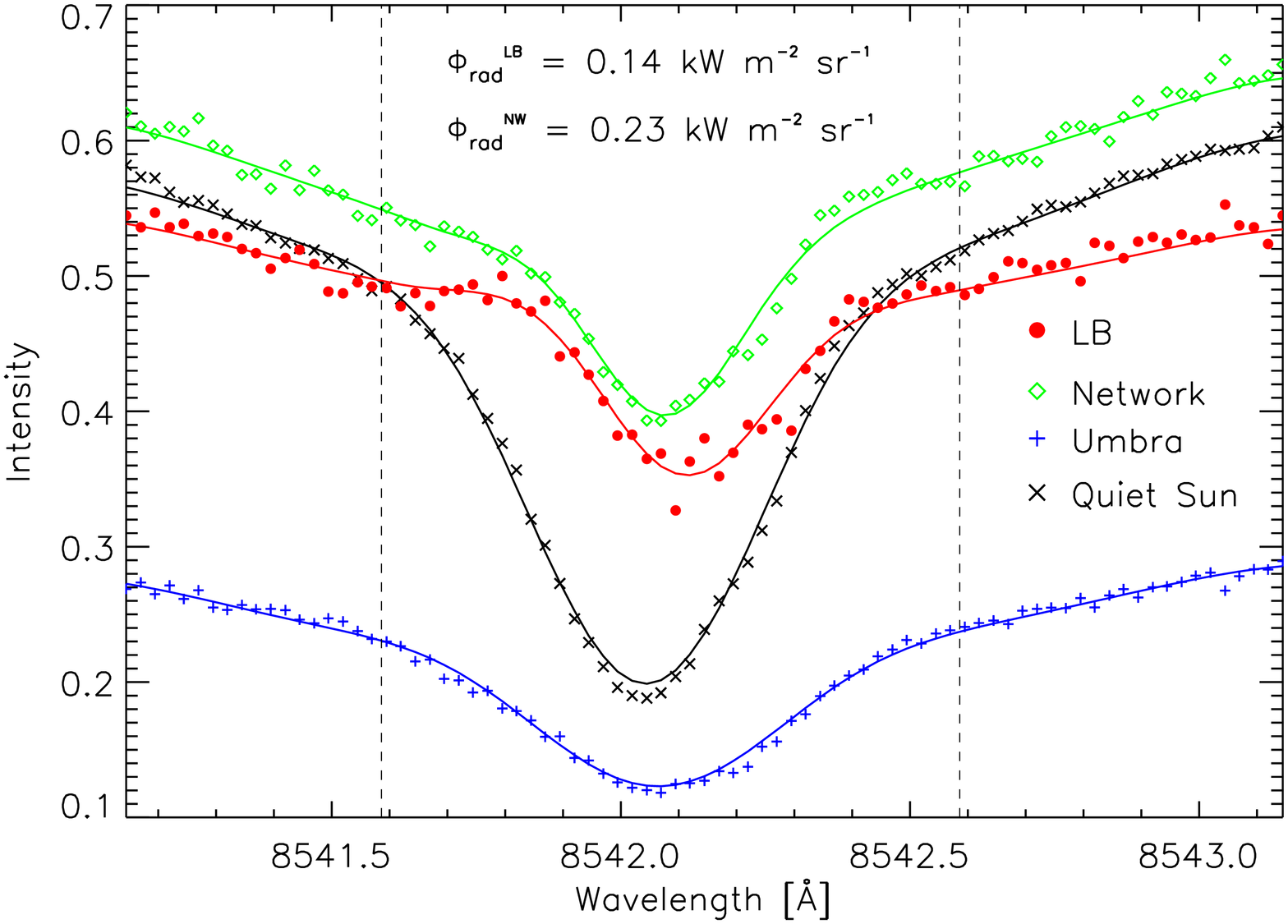}
}
\centerline{
\hspace{0pt}
\includegraphics[angle=90,width=\columnwidth]{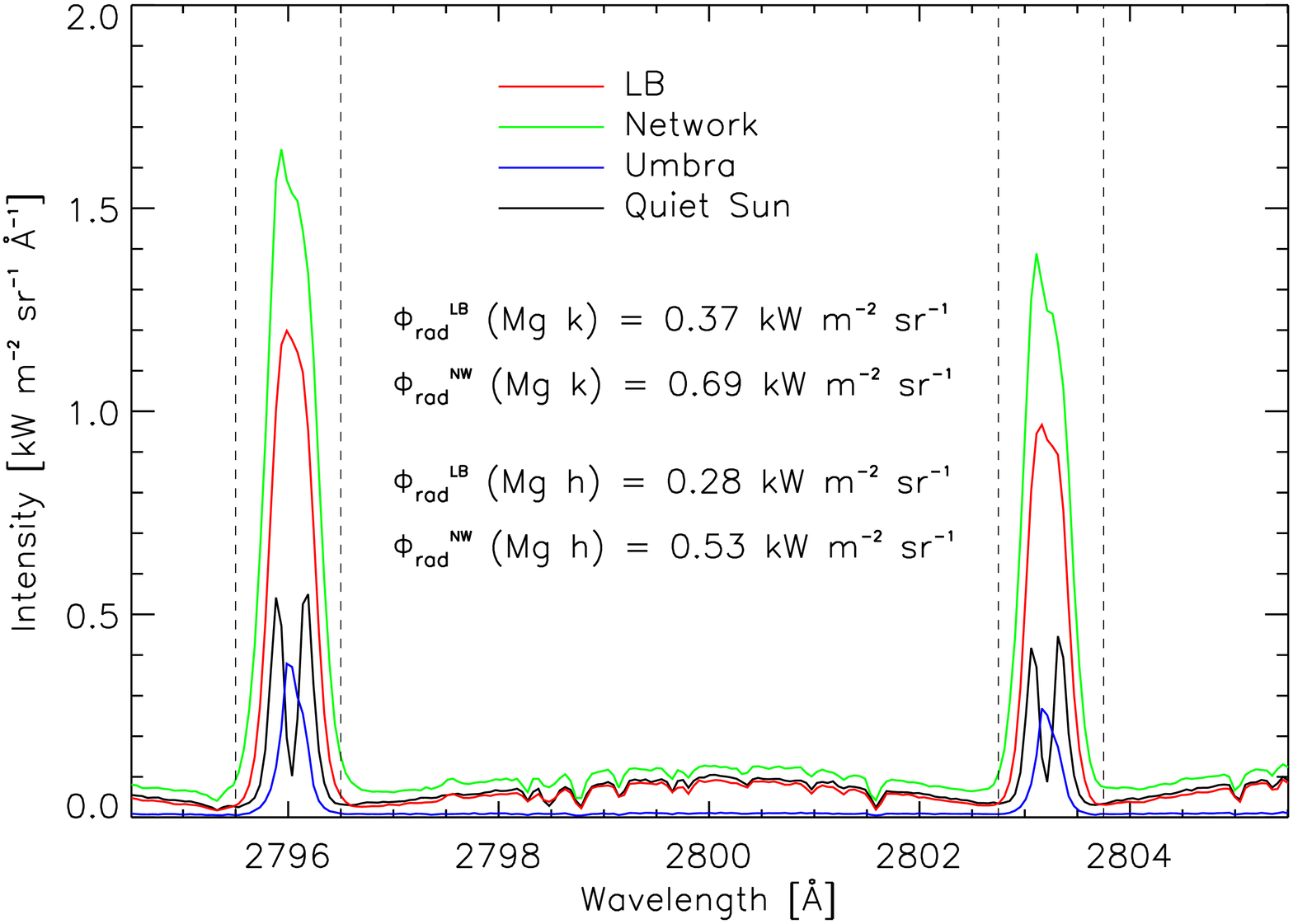}
}
\vspace{-5pt}
\caption{Top panel: MAST \ion{Ca}{2} spectra from different locations in the FOV. The 
symbols and the solid lines correspond to the observed and synthetic profiles, respectively. Bottom panel: 
observed IRIS \ion{Mg}{2} k \& h lines for the same locations. The dashed vertical lines mark the wavelength 
region within which the excess radiative losses were calculated. Their values for the network (NW) and the LB are 
given inside the panels.}
\label{fig14}
\end{figure}

\noindent{}where $A$ is the area of the LB, and $\rho$ is the coronal gas density. 
The loop height $h$ as derived from the extrapolations is about 12.3\,Mm.
Using values of $\rho$ of $5\times10^{-11}$\,kg\,m$^{-3}$, $g_\odot$ as 273.7\,m\,s$^{-2}$ 
and $A$ of 25.8\,Mm$^2$ and 35.7\,Mm$^2$ on May 14 and 15, respectively, $E_{fall}$ is 
estimated to be about $6.3\times10^{26}$\,erg and $8.7\times10^{26}$\,erg on May 14 and 15, respectively; 

The energy estimates from the different physical mechanisms that could be the sources are 
summarized in Table~\ref{tab02}.


\subsection{Estimates of Radiative Losses}
\label{rad_loss}

The top panel of Figure~\ref{fig14} shows \ion{Ca}{2} spectra and their synthetic fits from different 
regions in the FOV, i.e., the umbra, LB, QS, and magnetic network. The excess radiative loss in the LB with 
respect to the QS is about 0.14\,kW\,m$^{-2}$\,ster$^{-1}$ as derived from the calculation described 
in Sect.~\ref{radcalc}. In comparison, the excess loss in the network region is about 1.6 times higher 
at 0.23\,kW\,m$^{-2}$\,ster$^{-1}$. The excess radiative losses in the LB over the QS as estimated from 
the \ion{Mg}{2} k \& h lines are 0.37\,kW\,m$^{-2}$\,ster$^{-1}$ and 0.28\,kW\,m$^{-2}$\,ster$^{-1}$, 
respectively (bottom panel of Figure~\ref{fig14}). The spectral synthesis of the LB and QS temperature 
stratifications yields excess radiative losses in the LB that are a factor of 2--3 smaller than for 
the observations (second row in Table \ref{tab03}) with, e.g., 0.1\,kW\,m$^{-2}$\,ster$^{-1}$ for 
\ion{Mg}{2} h and 0.09\,kW\,m$^{-2}$\,ster$^{-1}$ for \ion{Ca}{2} IR at 854\,nm. The difference to 
the observations is presumably caused by the assumed density stratification in the synthesis and 
the lack of 3D radiative transfer. The direct observations can be taken to be more accurate in this 
context. 

\begin{table*}[!ht]
\caption{Estimates of radiative losses in different phenomena in excess of the QS. 
\ion{Ca}{2} IR refers to the spectral line at 854.2\,nm.} 
\label{tab03}
\begin{center}
\hspace{-50pt}
\begin{tabular}{c|l|c|c|ccccccc}
\hline
\multirow{2}{*}{No.} & \multirow{2}{*}{Type} & \multirow{2}{*}{$\Theta$ ($^\circ$)} & \multirow{2}{*}{Reference} &  \multicolumn{7}{c}{Excess Radiative Loss $\Phi_{rad}$ (kW\,m$^{-2}$\,ster$^{-1}$)} \\ \cline{5-11}
&  &  &  & \ion{Mg}{2} k & \ion{Mg}{2} h & \ion{Ca}{2} K & \ion{Ca}{2} H &  \ion{Ca}{2} IR & L$\alpha$ & H$\alpha$  \\
\hline \hline
\multirow{2}{*}{1.} & Sustained Heating in LB & \multirow{2}{*}{17} & \multirow{2}{*}{Current Study} & 0.37$^\textrm{\tiny{O}}$ & 0.28$^\textrm{\tiny{O}}$ & 0.27$^\textrm{\tiny{S}}$ & 0.19$^\textrm{\tiny{S}}$ & 0.14$^\textrm{\tiny{O}}$ & 0.03$^\textrm{\tiny{S}}$ & \\
 & Spectral Synthesis in LB  &    &               &   0.1  & 0.08 & 0.1 & 0.1 & 0.09 & 0.01 &  \\
\hline
2. & Polarity Inversion Line & 52 & Yadav et al. (2022) & 0.28 & 0.25 & 0.6 & 0.43 & 0.31 & 0.07 \\
\hline
3. & Ohmic Dissipation in LB & 13 & Louis et al. (2021) & & & & & 0.35 & & \\
\hline
4. & Flare Ribbon   & 52 & Yadav et al. (2022) & 2.23 & 1.98 & 1.76 & 1.34 & 0.78 & 0.57 \\
5. & Flare Ribbon   & 29 & IBIS DST 2014/10/24& & & & & 0.89 & & 2.17\\ 
\hline
6. & Ellerman Bomb  & 70 & Rezaei \& Beck (2015) & & & & 11 & 1.5 & & 4 \\
\hline
\end{tabular}
$^{\rm O}$: derived from observations $ $ $ $ $^{\rm S}$: derived from scaling factors from the second row 
\end{center}
\end{table*}

The factors for the radiative loss in the LB over the QS for the \ion{Ca}{2} K \& H lines and Ly$\alpha$ 
are 0.65, 0.46, and 0.08 times the \ion{Ca}{2} IR triplet (3$\times$\ion{Ca}{2} IR at 854\,nm), respectively, 
using the values from the third row of Table~\ref{tab03}. Adding the contributions from all the spectral lines 
in the first row of Table~\ref{tab03} and integrating over the solid angle of $4\pi$, the total chromospheric 
radiative loss in the LB is about 19.7\,kW\,m$^{-2}$ in excess of the QS. In terms of energy the above value 
translates into 3$\times$10$^{26}$\,erg using the LB area of 25.8\,Mm$^2$ and a time scale of 1\,min, where 
the latter takes the chromospheric relaxation time \citep{2008A&A...479..213B} into account, given that the 
heating in the LB is persistent over days and thus needs to be replenished continuously. For comparison, the 
total thermal energy in the EUV (1.9$\times$10$^{26}$\,erg), UV (8.6$\times$10$^{25}$\,erg), and visible 
(4.7$\times$10$^{25}$\,erg) on May 14 (refer Table~\ref{tab02}), is about 3.2$\times$10$^{26}$\,erg together, 
which gives a close match to the energy in the radiative losses that are a sink of energy.

\begin{figure}[!h]
\centerline{
\hspace{0pt}
\includegraphics[angle=90,width=\columnwidth]{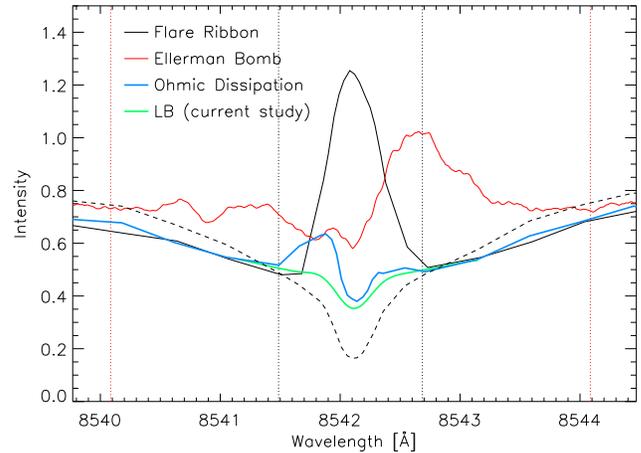}
}
\vspace{-5pt}
\caption{Illustrative spectra of different phenomena (solid lines). The black dashed 
line represents the QS profile. The vertical red dotted lines at $\pm 2$\,{\AA}\, are the wavelength 
regions within which the radiative loss for the EB was calculated, while the vertical black 
dotted lines correspond to the range used for the flare ribbon and ohmic dissipation.}
\label{fig15}
\end{figure}

Figure~\ref{fig15} shows representative spectra of the \ion{Ca}{2} 854.2\,nm line from a few other phenomena 
such as an Ellerman bomb (EB), a flare ribbon, and a case of ohmic heating in an LB for comparison, whose radiative 
losses are also listed in Table~\ref{tab03}. The radiative loss in an EB is about 
1.5\,kW\,m$^{-2}$\,ster$^{-1}$ \citep{2015A&A...582A.104R}, while that of a flare ribbon is about 
0.89\,kW\,m$^{-2}$\,ster$^{-1}$. Similarly, ohmic dissipation in an LB comprises about 0.35\,kW\,m$^{-2}$\,ster$^{-1}$ 
\citep{2021A&A...652L...4L}, which is about 2.5 times greater than the value obtained for the LB under 
investigation. The current case of continuous long-term heating is thus at the lower end of the energy range 
of more short-lived and dynamic events.

\section{Discussion}
\label{discuss}
\subsection{LB Properties and Evolution}
The granular LB that formed in a regular, unipolar sunspot remained stable until the AR traversed the solar 
limb. During this time, the host sunspot did not fragment nor were there any large-scale changes in the 
magnetic structure of the AR. Measurable changes were seen in the flux of the sunspot and the area of 
the LB, which decreased and increased by $1.7\times10^{21}$\,Mx and 22\,Mm$^2$, respectively, 
over the course of four days. While overturning convection is prevalent in the LB at photospheric heights, 
the LB is heated over a large temperature range from 8000\,K to 2.5\,MK spanning the chromosphere to
the low corona that is sustained for more than two days. 
The signatures of this heating are seen in the temperature maps of the chromospheric \ion{Ca}{2} and 
\ion{Mg}{2} spectral lines, the peak amplitude and line width of the IRIS \ion{C}{2}, \ion{Si}{4} lines 
that form at a temperature of 30,000\,K and 65,000\,K, respectively, and the emission measure using the 
different AIA channels. The enhanced intensities and line widths of the IRIS lines in the LB 
are in good agreement with \cite{2018A&A...609A..73R}.
The persistent heating over the LB is counterintuitive as the underlying structure would radiate the 
majority, if not all, of its energy once having evolved to a strongly convective region inside the sunspot. 
We now discuss the possible mechanisms that can provide the necessary energy to sustain the enhanced temperature 
over the LB for more than two days.

\subsection{LB Energy Budget}

\paragraph{Thermal Energy and Radiative Losses}
An estimate of the thermal energy in the visible, NUV, and EUV yields about $3.2\times10^{26}$\,erg 
using the values on May 14 as shown in Table~\ref{tab02}. The chromospheric temperature in the LB is 
enhanced compared to its immediate umbral surroundings and even with respect to QS conditions. The enhanced 
temperature persists over a few days in a similar way. The heating process has thus lifted the temperature 
to a new, higher energetic equilibrium that is maintained in time.  

The estimate of the chromospheric instantaneous excess energy losses over the LB area yields 
$3\times10^{26}$\,erg, which gives a close match to the predicted losses from the temperature excess. 
This confirms the presence of a new stationary equilibrium at a higher energy level than, e.g, in the QS. 
In comparison to typical values for other chromospheric heating events such as ohmic dissipation 
\citep{2021A&A...652L...4L}, flare ribbons \citep{2022A&A...665A..50Y} or an Ellerman bomb 
\citep{2015A&A...582A.104R} the current radiative losses are found to be at the low end of the range. 
Apart from ohmic dissipation, the other types of heating events are generally impulsive and short-lived 
with time scales of only a few tens of minutes with spatially localized heating sources from reconnection 
\citep{georgoulis+etal2002} or particle beams \citep{2016ApJ...816...88K}. While flare ribbons can cover 
similar areas as the current LB, the heating process that causes its enhancement must be both spatially 
extended and long-lived, albeit at a 6--10 times lower heating rate than for the more impulsive events.

\paragraph{LB Expansion and Freefall Acceleration}
The kinetic energy associated with the LB expansion is 1--2 orders of magnitude higher than 
the net thermal energy. This mechanism is related to photospheric dynamics that can lead to the buildup 
of energy in the corona. \cite{2011ApJ...729...97M} showed that photospheric footpoint motions in a decaying 
AR could store enough free magnetic energy in the corona to compensate the radiative losses. 
Using idealized numerical models, \cite{2002ApJ...577..993H} showed that by directly coupling 
compressible magneto-convection at the bottom layer to a low plasma-$\beta$ region above, the overlying corona 
could be heated by the Poynting flux emerging from the upper boundary.
The LB expansion continues on the same temporal and spatial scales as the heating process over days. The 
heating seems to be localized above the LB area also in the higher atmosphere, which indicates a correlation 
to the photospheric spatial pattern. It is unclear, however, how the photospheric mechanical energy would be 
transported upward and deposited locally in the chromosphere and transition region in the current case.

Freefall acceleration of plasma along loops \citep{2021ApJ...916....5S}
connecting the opposite polarity network flux to the LB or close to it in the sunspot could also provide 
sufficient energy to sustain the temperature over the LB. The long-lived presence of loops at or near the 
LB and the persistent brightness in the AIA channels suggest that this could also be a likely possibility. 
However, with the exception of a few locations in the umbra next to the LB we do not see any strong redshifts 
in the \ion{C}{2} or \ion{Si}{2} lines that would indicate plasma draining down the loop from a height of 12\,Mm. 
Coronal rain was also found to be often rather intermittent in nature without continuous flows 
\citep[][]{antolin+etal2012,li+etal2021}.

Furthermore, we do not unambiguously detect blueshifts at the other end of the loops, which would provide 
evidence for a siphon flow \citep{1980SoPh...65..251C,2015A&A...582A.116S,2022A&A...662A..25P}. 
The LB along with the network flux region are redshifted, which raises questions on how any flow would be 
sustained for a period of days. While the Doppler shift measurements in the LB do not reveal high-speed
downflows all the time, the enhanced intensity as well as line width are maintained in the LB for over two days.

\paragraph{Magnetic Flux Losses and Field-related Heating}
The energy corresponding to the magnetic flux loss of the sunspot or the apparent gain of magnetic flux above the 
LB does also match the energy requirements from the radiative losses and the thermal energy. They occur continuously 
on the same time scale as the persistent LB heating. The total loss of magnetic flux of the sunspot would, however, 
not match the required localized heating with a preferred occurrence above the LB area, while the apparent gain of 
magnetic flux above the LB area would require a nearly complete conversion of magnetic to thermal energy to balance 
the radiative losses. \cite{2012ApJ...759..104H} showed that magnetic flux dispersal at the photosphere is important 
for the release of nonthermal energy in the corona for a decaying AR that was devoid of flux emergence. 
The photospheric interactions of a bipole containing a flux of $2\times 10^{18}$\,Mx with an overlying 
field via cancellation, emergence, and relative photospheric motions could dissipate about $1.3$--$3.2\times 10^{26}$\,erg 
in the corona over a time interval of 100 minutes \citep{2012SoPh..278..149M}.
Using the magnetofrictional approach of \cite{1986ApJ...309..383Y}, 
\cite{2013ApJ...770L..18M} evolved the corona through a sequence of nonpotential, quasi-static equilibria using 
photospheric LOS magnetograms at the bottom boundary. They found that the storage of energy and its subsequent 
dissipation in the quiet corona occurred at a mean rate of $8.7\times 10^{4}$\,erg\,cm$^{-2}$\,s$^{-1}$, and produced 
dark and bright features similar to those in EUV images.
The so-called braiding of magnetic flux by random photospheric motions \citep{parker1972,peter+etal2004} could set in 
at the boundary layer between the presumably field-free overturning convection in the LB and the surrounding umbral 
magnetic fields. For the case of the current LB, the resulting heating would, however, have to set in very low in 
the atmosphere starting at chromospheric levels.

\paragraph{Wave Heating} 
As the LB exhibits convective motions that are possibly rooted 
quite deep, magneto-acoustic waves could dissipate a part of their energy in the higher atmospheric layers 
\citep{1978A&A....70..487U,2007ApJ...671.2154K,2012ApJ...746...68K,2018MNRAS.479.5512K}. 
The acoustic energy flux estimated from a number of chromospheric lines shows that at least in the quiet Sun, 
it is able to balance the radiative losses at heights between 900 and 2200\,km 
\citep{2020ApJ...890...22A,2020A&A...642A..52A}. However, in active regions the acoustic flux 
balances only  10--30\% of the radiative losses \citep{2021A&A...648A..28A}. This is also in agreement with 
previous studies by \cite{2009A&A...507..453B}. Based on the above, and coupled with the lack of time series 
IRIS observations with high temporal cadence, one can argue
that the residual acoustic flux would only have a minor, if not negligible, contribution in heating the LB 
to transition region and coronal temperatures. On the other hand, Alfv\'en waves have been proposed as 
possible energy transporters that could heat the upper atmospheric layers 
\citep{1961ApJ...134..347O,1981ApJ...246..966S,2011ApJ...736....3V,2020ApJ...900..120S} with direct evidence 
for an energy deposit in the chromosphere in \cite{2018NatPh..14..480G}. Thus, we cannot rule out the possibility  
of Alfv\'en waves heating the chromosphere and transition region above the LB, with the caveat that the nearly 
vertical smooth magnetic field would make a mode conversion and subsequent energy deposit difficult.

\paragraph{Ohmic Dissipation}
Ohmic dissipation in the LB could arise from electric currents due to the presence of weak 
magnetic fields inside the sunspot. Recently, \cite{2020ApJ...905..153L}
reported the bodily emergence of horizontal magnetic fields along an LB that 
comprised strong blueshifts all along the LB lasting for a period of 13\,hr. The emergence of 
flux rendered strong electric currents leading to ohmic dissipation that was accompanied by 
large temperature enhancements in the chromosphere above the LB \citep{2021A&A...652L...4L}.  
A similar observation of blueshifts and chromospheric
emission was seen during the emergence of a small-scale, bipolar loop in a granular LB 
\citep{2015A&A...584A...1L}. However, the granular LB analyzed here did not comprise any strong 
or significant velocities in the photosphere, and the photospheric currents were very weak or negligible.
For ohmic dissipation to play a role in the chromosphere and above, the currents at the bottom boundary 
have to be the strongest for them to be significant in the higher layers, which is not the case here. 

\paragraph{Transient Chromospheric Events}
LBs are known to exhibit a range of transient phenomena, including jets \citep{2014A&A...567A..96L}, 
brightenings \citep{2008SoPh..252...43L}, and flares \citep{2021ApJ...907L...4L}. There were several confined, 
however weak, flares originating in the AR on May 15 and an eruptive flare on May 13 which resulted in one of 
the flare ribbons, although compact, to be co-spatial with the LB. However, there were no flares associated 
with the LB or the AR on May 14. The spatial coincidence of one of the flare ribbons along the LB and the 
ensuing post flare loops extending to the LB indicate the connectivity of the LB to the large-scale topology 
of the AR, which was destabilized by the erupting filament below. The association of an LB with the large-scale 
topology of an AR has also been observed by \cite{2010ApJ...711.1057G}, where repetitive surges from an LB  led 
to the eruption of an adjacent filament. While the flares could play an important role in depositing energy in 
the higher layers of the LB, which would subsequently heat the lower layers, the strength of the flares, and the 
rapid radiative cooling would not explain the persistent brightness of the LB over 48\,hr. In addition, the IRIS 
SJ images do not indicate any discernible, small-scale, reconnection-driven events during the raster scans, 
although we do not rule out the possibility that these could have been missed during the data gap. Even if 
there were small-scale ejections, they would be localized and would not explain the heating over the entire 
extent of the LB. 

The case of sustained heating over a granular LB is not uncommon. \cite{2003ApJ...589L.117B} reported a constant 
brightness enhancement over a granular LB using the 1600\,\AA\, channel of the Transition Region and Coronal 
Explorer \citep[TRACE;][]{1999SoPh..187..229H}. A C2.0 flare was also observed wherein one of the ribbons was 
co-spatial with the LB similar to the observations reported in this study. The authors attributed the persistent 
brightness to the stressed magnetic configuration at the LB that could lead to reconnection and energy dissipation. 
As stated above, the lack of electric currents rules out ohmic dissipation as the source(s) of heating over the LB.   

\subsection{Possible Heating Process(es)}
The energy estimates associated with the loss/gain of magnetic flux, increase in the LB area, 
and freefall acceleration exceed the thermal energy in the LB that matches the radiative 
losses. However, it remains unclear if one or a combination of the above processes are the 
primary source of heating over the LB. Only some of the possible processes match the necessary 
temporal (long duration) and spatial patterns (concentration on LB area). The heating rate is found to be lower 
than for other impulsive chromospheric heating events. A process related to the photospheric mechanical energy 
from either the LB expansion or the overturning convection inside the LB and its interaction with bordering 
magnetic fields seems to be more likely because a continuous driver over days is needed.

As pointed out by \cite{2018A&A...609A..73R}, LBs are multithermal structures, with diverse heating mechanisms 
supplying momentum and energy to different layers of the solar atmosphere. It remains an open question whether 
such a persistent heating over a large height range in a granular LB is indeed a generic phenomenon. 
In the current study, we could especially not determine whether the energy that heats the 
LB region at different heights is converted from other energy sources and deposited locally, or results from 
either an upward or downward energy transfer through the outer solar atmosphere.


\section{Conclusions}
\label{conclude}
The LB under investigation evolved sufficiently to exhibit overturning convection without fragmenting the 
regular, unipolar sunspot. Despite the absence of any large-scale flux emergence or apparent changes in the 
magnetic topology of the AR, the LB was associated with strong heating
spanning a temperature range of 8000\,K to 2.5\,MK, which was maintained for more than two days. In addition 
to the persistent brightness, large-scale coronal loops are always rooted at or close to the LB. The estimated 
thermal energy from the EUV, NUV, and visible spectral regions 
is about $3.2\times10^{26}$\,erg  and lines up with estimates of the chromospheric radiative 
losses. The continued heating could be accounted for by one, or a combination of the following processes, 
namely, loss of magnetic flux, kinetic energy from the lateral expansion of the LB 
or overturning convection inside it, and freefall acceleration of plasma along coronal 
loops. The absence of strong electric currents in the LB rules out heating by ohmic dissipation. Further studies 
are needed to determine if such sustained heating is a general characteristic of sunspot LBs  
or whether the LB in this study is a rare exception. 


\begin{acknowledgments}
{\footnotesize\textit{Acknowledgments. The 0.5~m Multi-Application Solar Telescope 
is operated by the Udaipur Solar Observatory, Physical Research Laboratory, 
Dept. of Space, Govt. of India. IRIS is a NASA small explorer mission 
developed and operated by LMSAL with mission operations executed at NASA 
Ames Research Center and major contributions to downlink communications 
funded by ESA and the Norwegian Space Centre. Hinode is a Japanese mission 
developed and launched by ISAS/JAXA, collaborating with NAOJ as a domestic 
partner, NASA and STFC (UK) as international partners. Scientific operation of 
the Hinode mission is conducted by the Hinode science team organized at 
ISAS/JAXA. This team mainly consists of scientists from institutes in 
the partner countries. Support for the post-launch operation is provided 
by JAXA and NAOJ(Japan), STFC (U.K.), NASA, ESA, and NSC (Norway).
The data in this article are courtesy of NASA/SDO and the AIA 
science team. SDO is a mission for NASA's Living With a Star (LWS) 
Program. Hinode SOT/SP Inversions were conducted at NCAR in the 
framework of the Community Spectro-polarimtetric Analysis Center 
(CSAC; http:\/\/www.csac.hao.ucar.edu\/). We thank Ms. Bireddy Ramya and
Ms. Anisha Kulhari at Udaipur Solar Observatory for assisting in the telescope 
operation and image acquisition. R.E.L. would like to thank
Graham Barnes at NWRA for providing the latest version of the AMBIG code
and Avijeet Prasad at the University of Oslo for carrying out the NFFF extrapolation. 
We thank J.~Jenkins for help with the synthesis. 
One of the authors Debi Prasad Choudhary was supported by the National Science Foundation 
with grant number AGS 2050340.
We would like to thank the referee for his/her suggestions.}}
\end{acknowledgments}

\bibliographystyle{aasjournal}
\bibliography{louis_ref}


\appendix
\label{app}

\section{MAST \ion{Ca}{2} and IRIS \ion{Mg}{2} Spectra} 

\begin{figure*}[!ht]
\centerline{
\hspace{75pt}
\includegraphics[angle=90,width=0.46\textwidth]{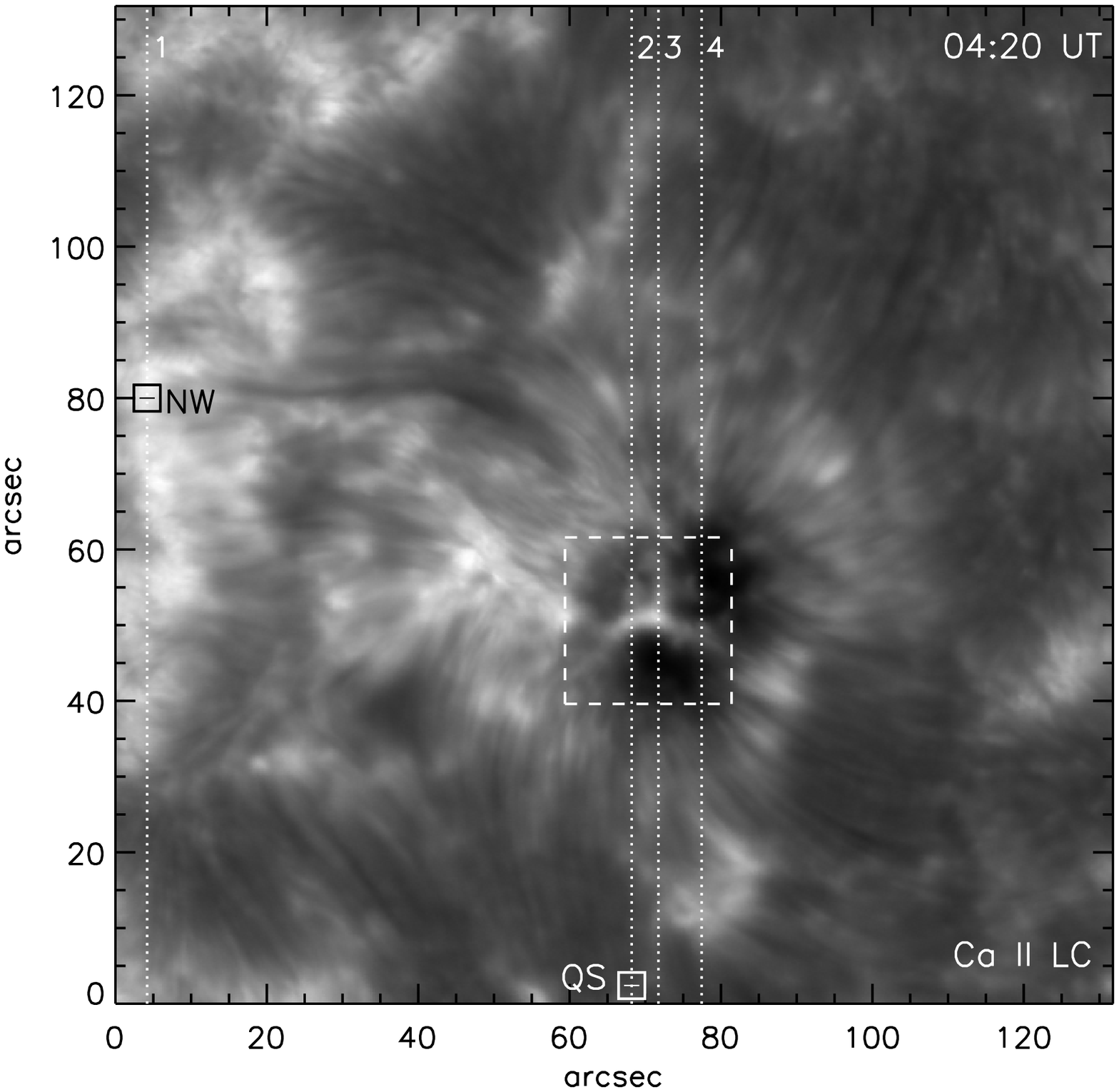}
\hspace{-80pt}
\includegraphics[angle=90,width=0.46\textwidth]{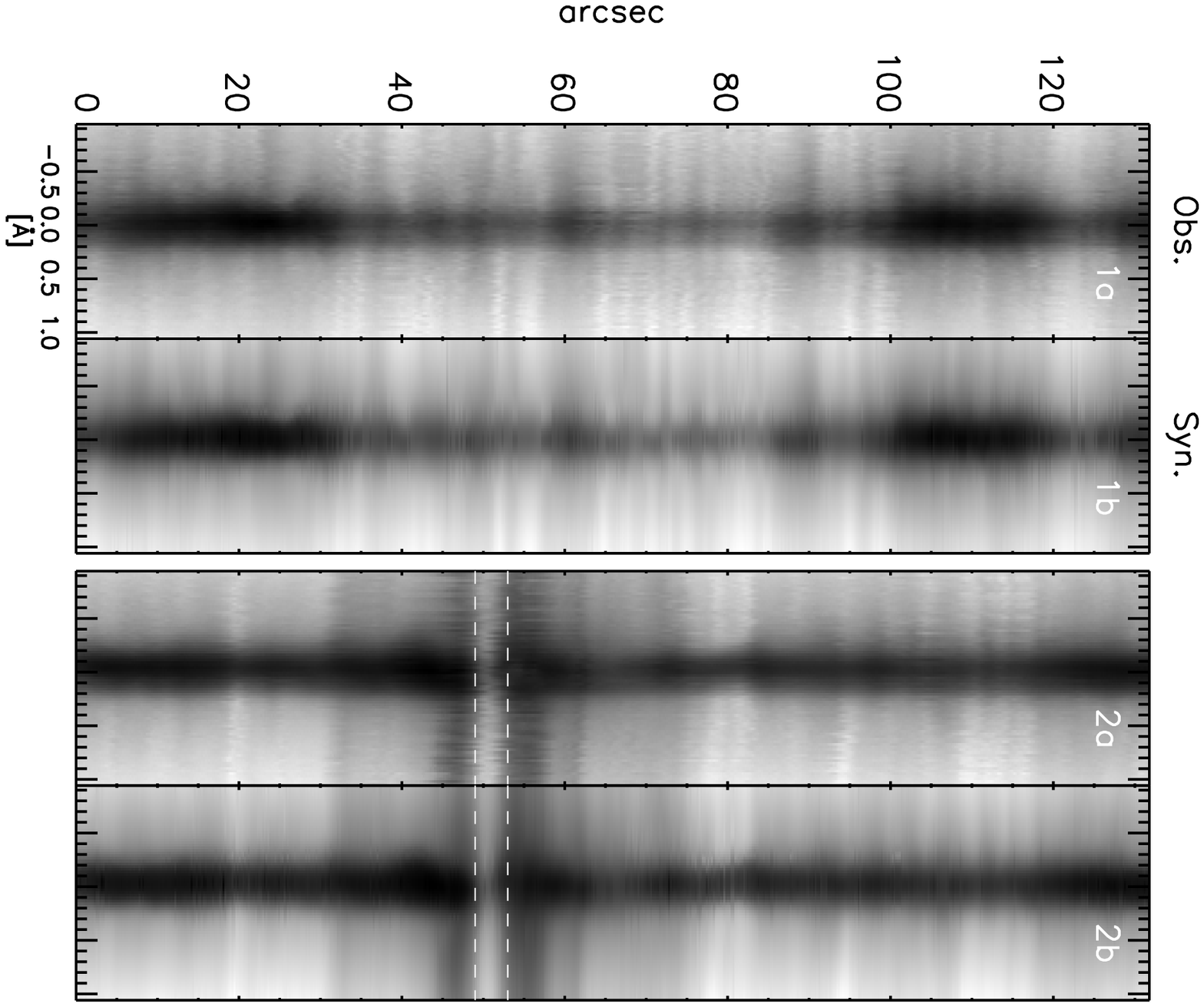}
\hspace{-124pt}
\includegraphics[angle=90,width=0.46\textwidth]{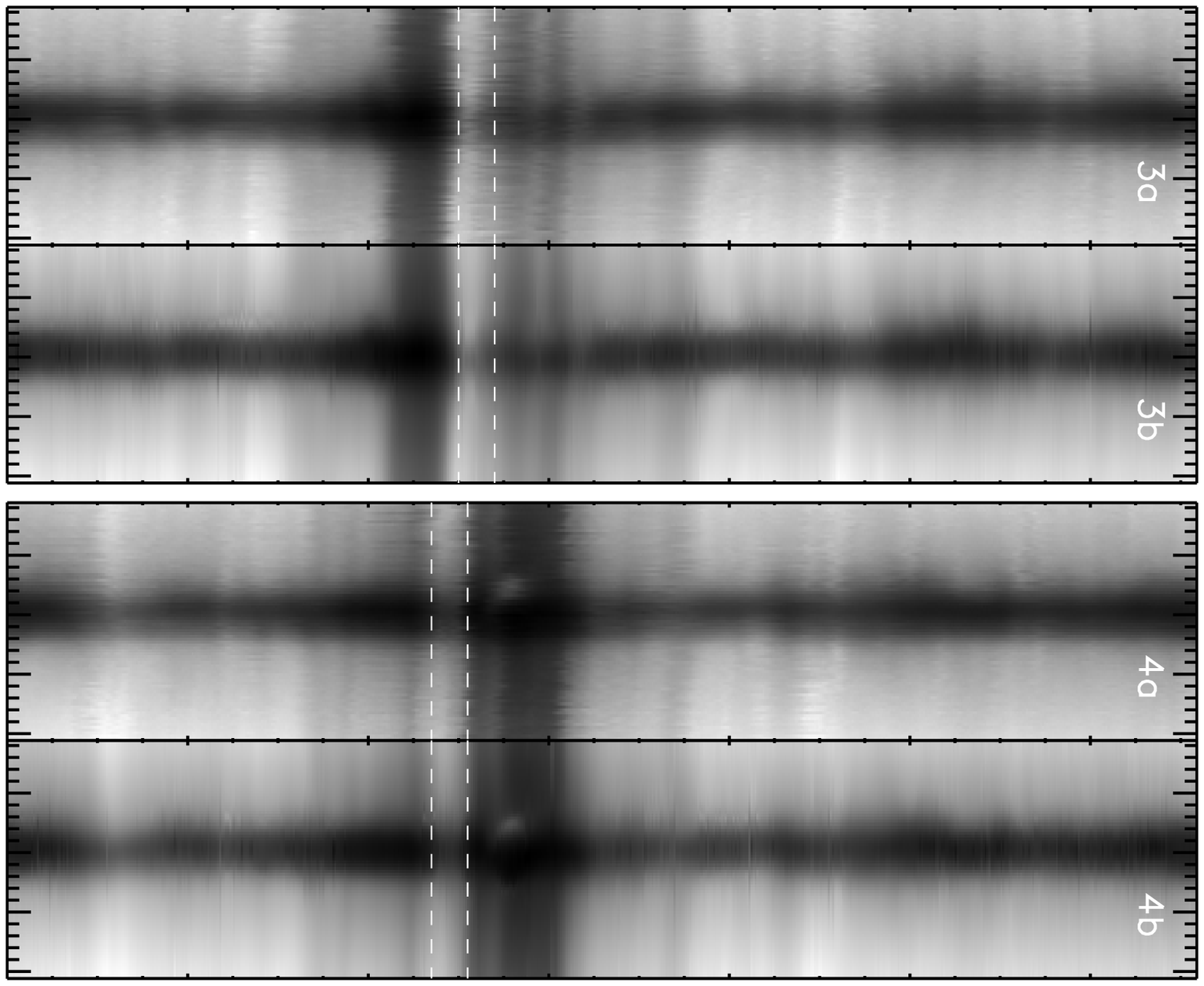}
\hspace{-124pt}
\includegraphics[angle=90,width=0.46\textwidth]{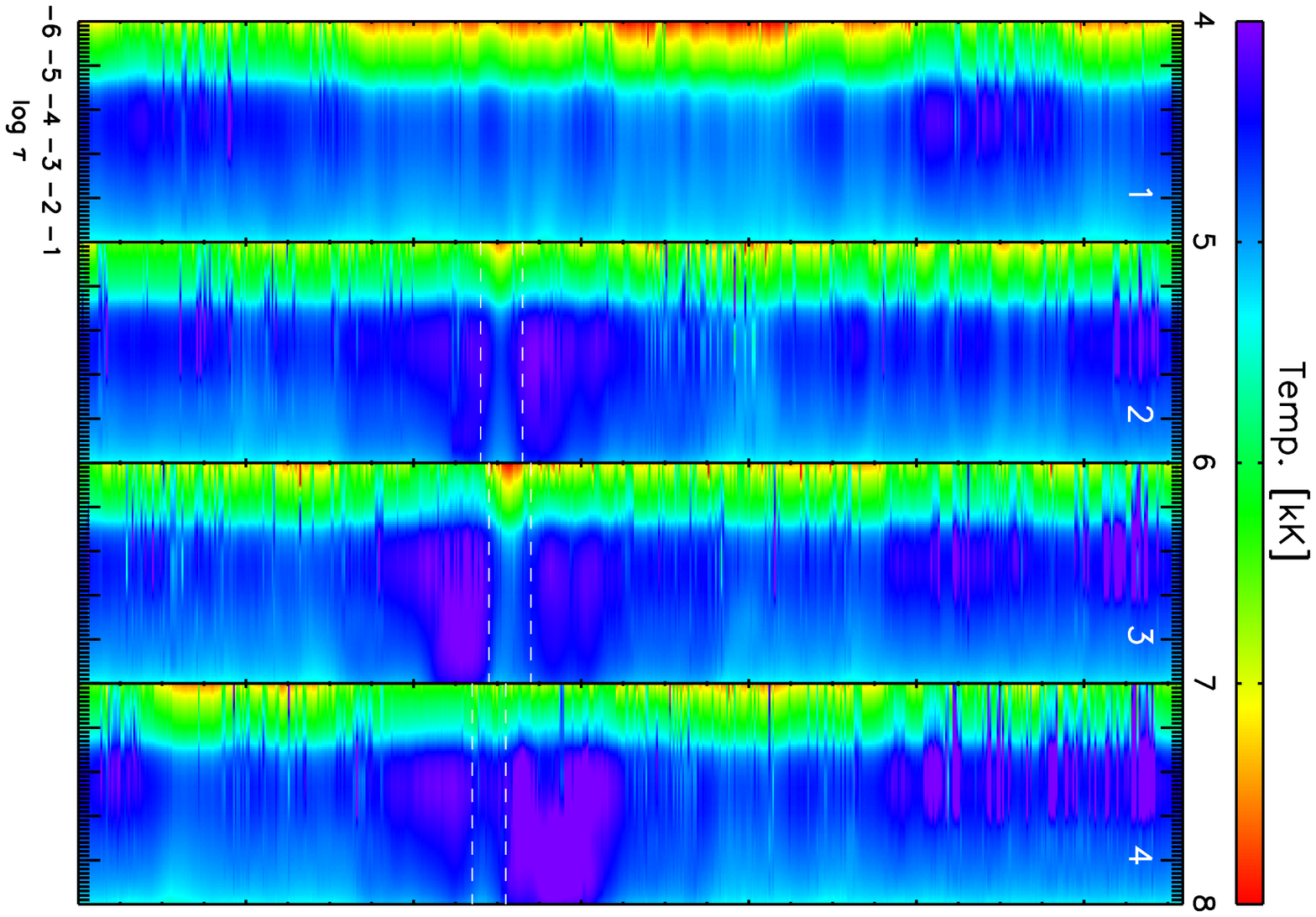}
}
\vspace{-10pt}
\centerline{
\hspace{75pt}
\includegraphics[angle=90,width=0.46\textwidth]{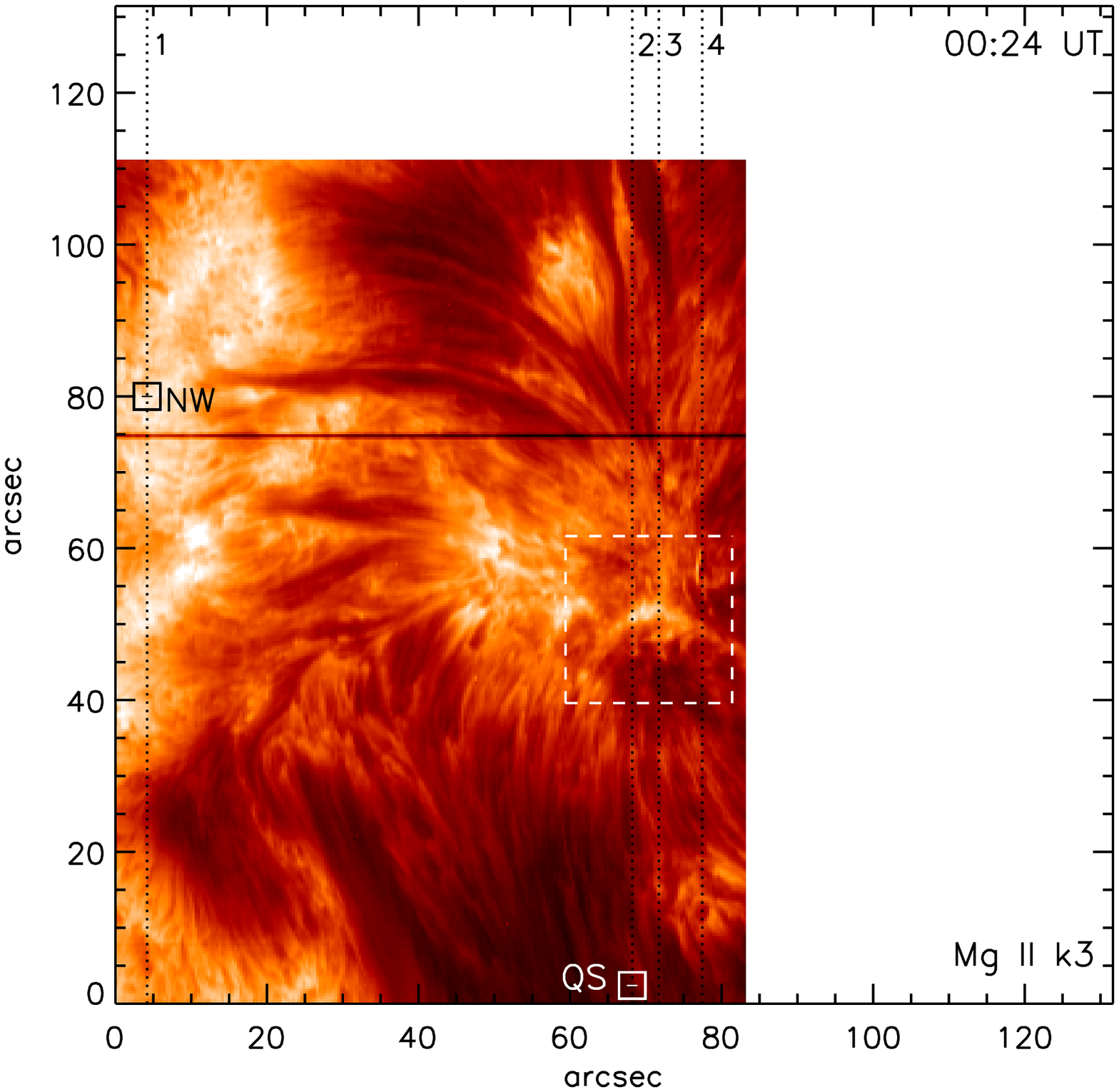}
\hspace{-80pt}
\includegraphics[angle=90,width=0.46\textwidth]{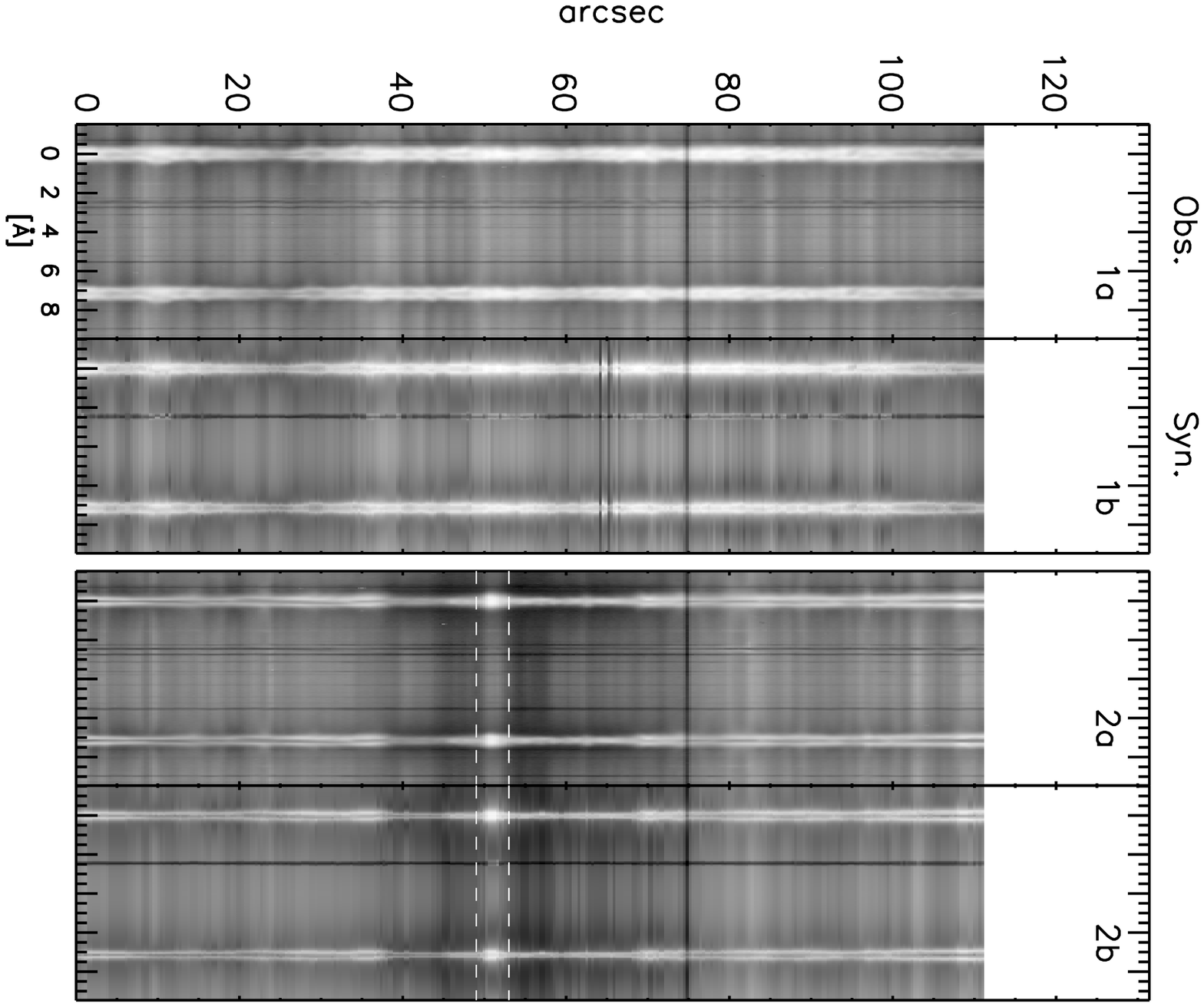}
\hspace{-124pt}
\includegraphics[angle=90,width=0.46\textwidth]{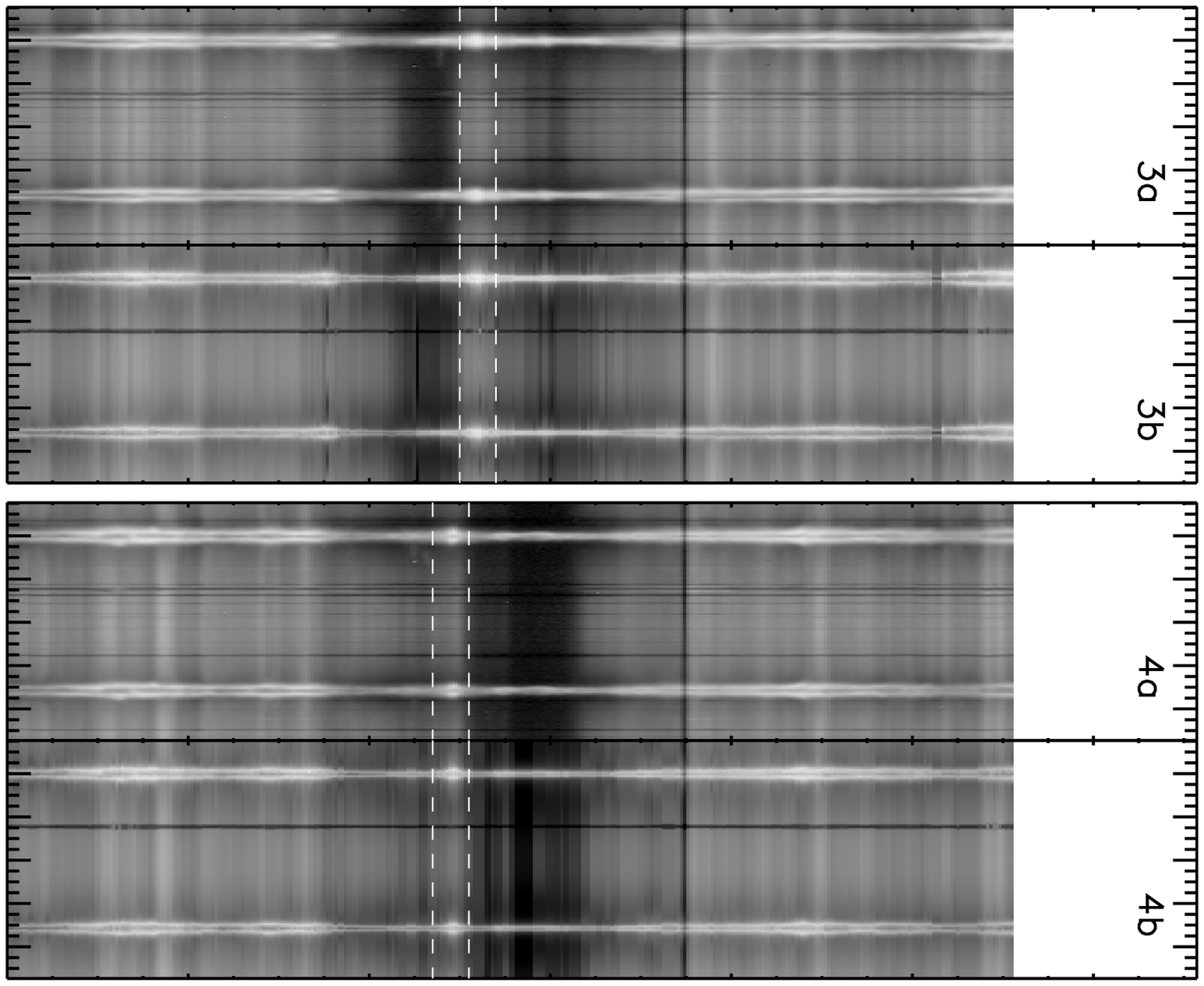}
\hspace{-124pt}
\includegraphics[angle=90,width=0.46\textwidth]{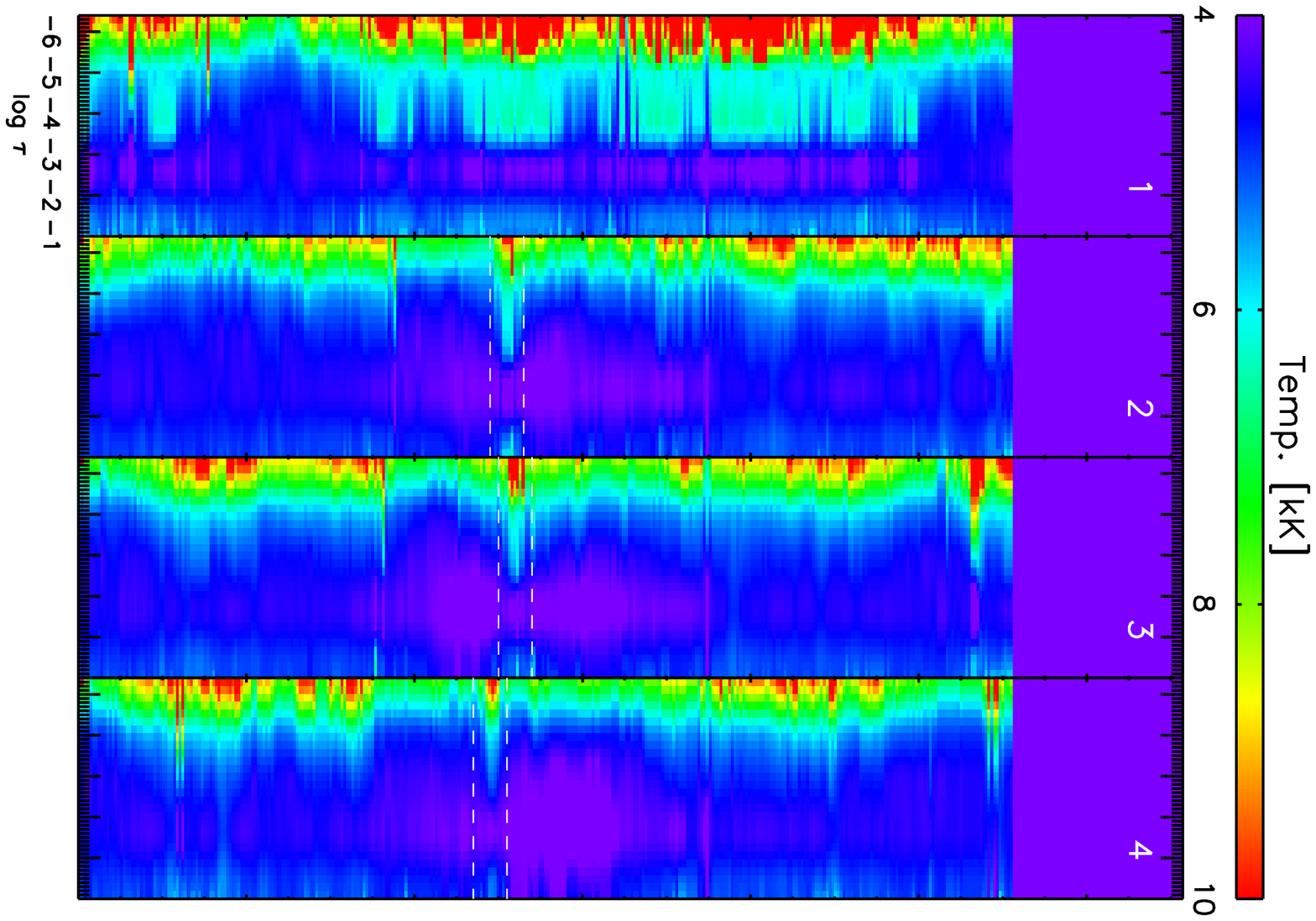}
}
\vspace{-5pt}
\caption{Top panels: NICOLE inversion results in the LB from MAST observations
of the \ion{Ca}{2} line at 854.2\,nm. The dashed white square indicates a smaller
FOV around the LB, which is depicted in Figure~\ref{fig04}. The locations marked `NW' and `QS'
correspond to the network and quiet Sun, respectively, and whose profiles are shown in Figure~\ref{fig14}.
The middle panels show the observed 
and synthetic spectra for the four vertical cuts in the line core image. The right panels
correspond to the temperature stratification along the four cuts. 
The white horizontal dashed lines in the middle and right panels depict the spectral region 
corresponding to the LB and the associated temperature stratification, respectively.
Bottom panels: Same as above but for the IRIS \ion{Mg}{2} line.}
\label{app_fig01}
\end{figure*}

Figure~\ref{app_fig01} shows the observed and synthetic spectra in the MAST \ion{Ca}{2} 
and IRIS \ion{Mg}{2} lines along with their temperature stratifications for a few spatial 
locations in the FOV. The synthetic spectra from the NICOLE and IRIS$^2$ inversions match 
the observed spectra quite well with the resulting temperature stratification having a 
nearly smooth variation in the spatial and height domain. While the \ion{Ca}{2} and 
\ion{Mg}{2} lines both show an enhancement in temperature over the LB at heights above 
$\log\tau=-3$, the spectral signatures in the two lines are quite distinct. 

At the highest layers near $\log\tau=-6$, the temperature in the LB is nearly comparable 
to that in the opposite polarity network flux region being cooler than the latter by about 
100\,K\, as estimated from the \ion{Ca}{2} line. In comparison, the temperature difference 
is about 800\,K\, in the \ion{Mg}{2}. However, the stratification in the lower
heights is very similar between the LB and the network region as seen in both lines. 

\section{IRIS NUV and FUV lines}

\begin{figure*}[!ht]
\centerline{
\hspace{75pt}
\includegraphics[angle=90,width=0.65\textwidth]{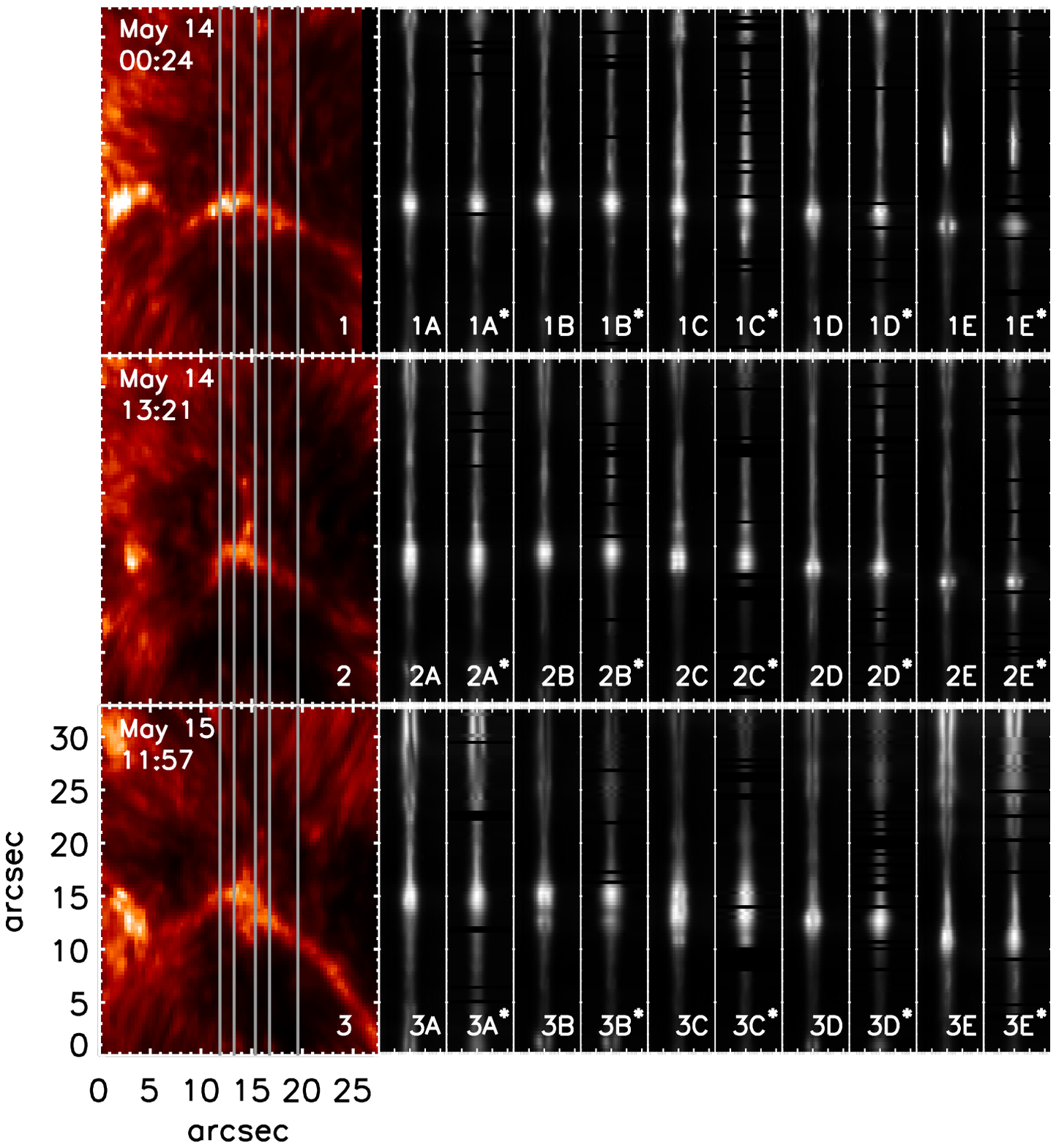}
\hspace{-180pt}
\includegraphics[angle=90,width=0.65\textwidth]{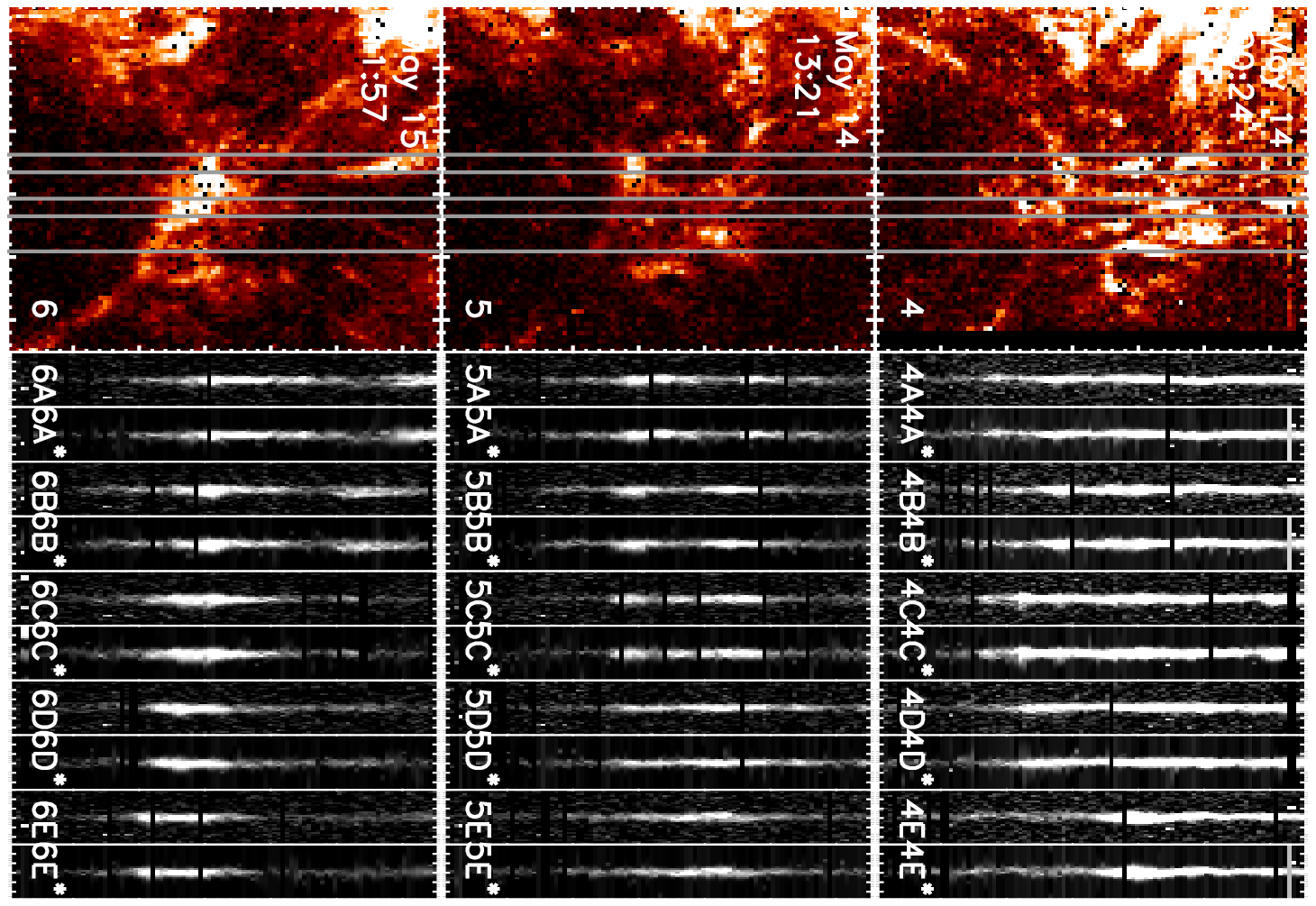}
\hspace{-219pt}
\includegraphics[angle=90,width=0.65\textwidth]{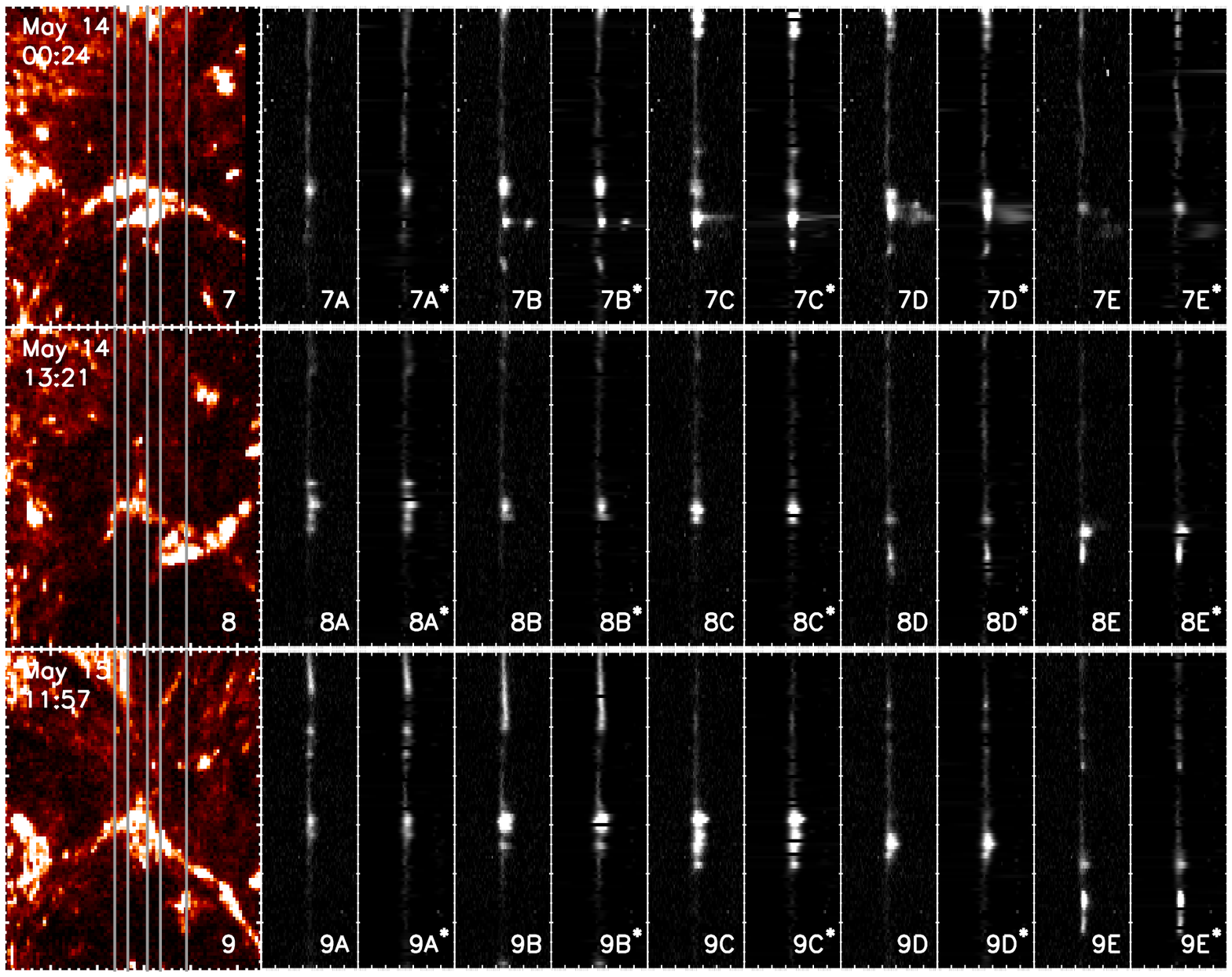}
}
\vspace{-50pt}
\caption{Spectral profiles of IRIS lines and their corresponding fits. Panels 1, 4, and 7 correspond
to the maximum intensity of the \ion{Mg}{2} k line, the \ion{C}{2} line, and \ion{Si}{4}
line, respectively. The vertical lines represent cuts whose spectra are shown on the right of each panel 
indexed A--E, while the corresponding fits are indicated with an $\ast$. The fits to spectra were obtained 
using a double Gaussian profile.}
\label{app_fig02}
\end{figure*}

Figure~\ref{app_fig02} shows the peak intensity and spectra in the IRIS NUV and FUV lines in the LB 
at different instances of time spanning a duration of 36\,hr. The panels to the right of the LB FOV 
correspond to the observed and fitted spectra with the latter being derived from a double Gaussian 
fit and indexed with an $\ast$. All three IRIS lines show the LB to have a pronounced line width 
although this is less conspicuous for the \ion{C}{2} on May 14. While the \ion{Mg}{2} and \ion{C}{2}
lines show the LB with an enhanced intensity, the \ion{Si}{4} lines show additional structures that 
are possibly related to the coronal loops ending in the sunspot and are often seen very close to the LB.
These secondary structures sometimes show redshifts in excess of 100\,km\,s$^{-1}$\, which are reproduced
very well in the double Gaussian fit. 

\section{Loop Height Estimate from AIA Images}

\begin{figure*}[!ht]
\centerline{
\hspace{100pt}
\includegraphics[angle=90,width=0.7\textwidth]{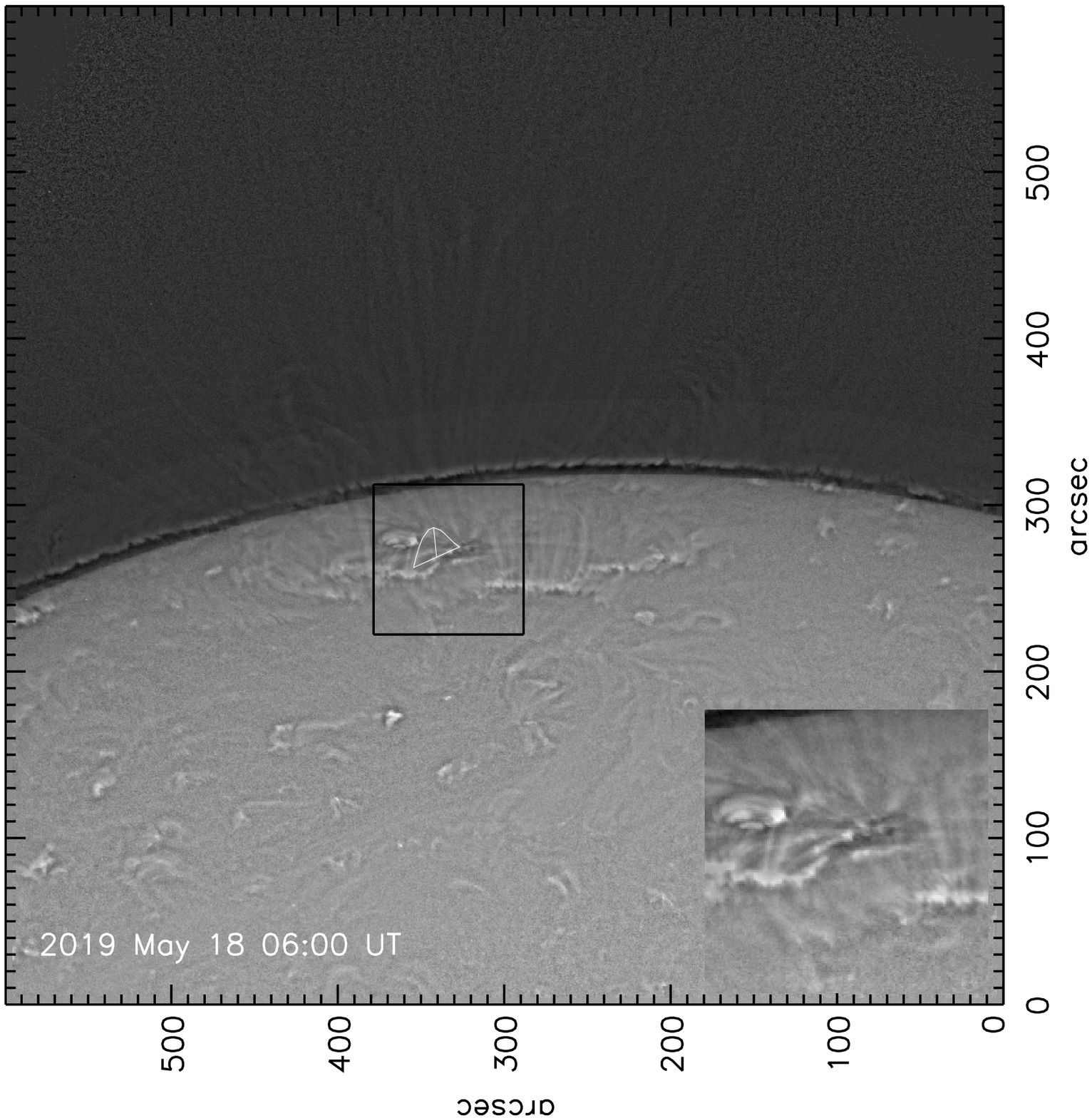}
\hspace{-100pt}
\includegraphics[angle=90,width=0.7\textwidth]{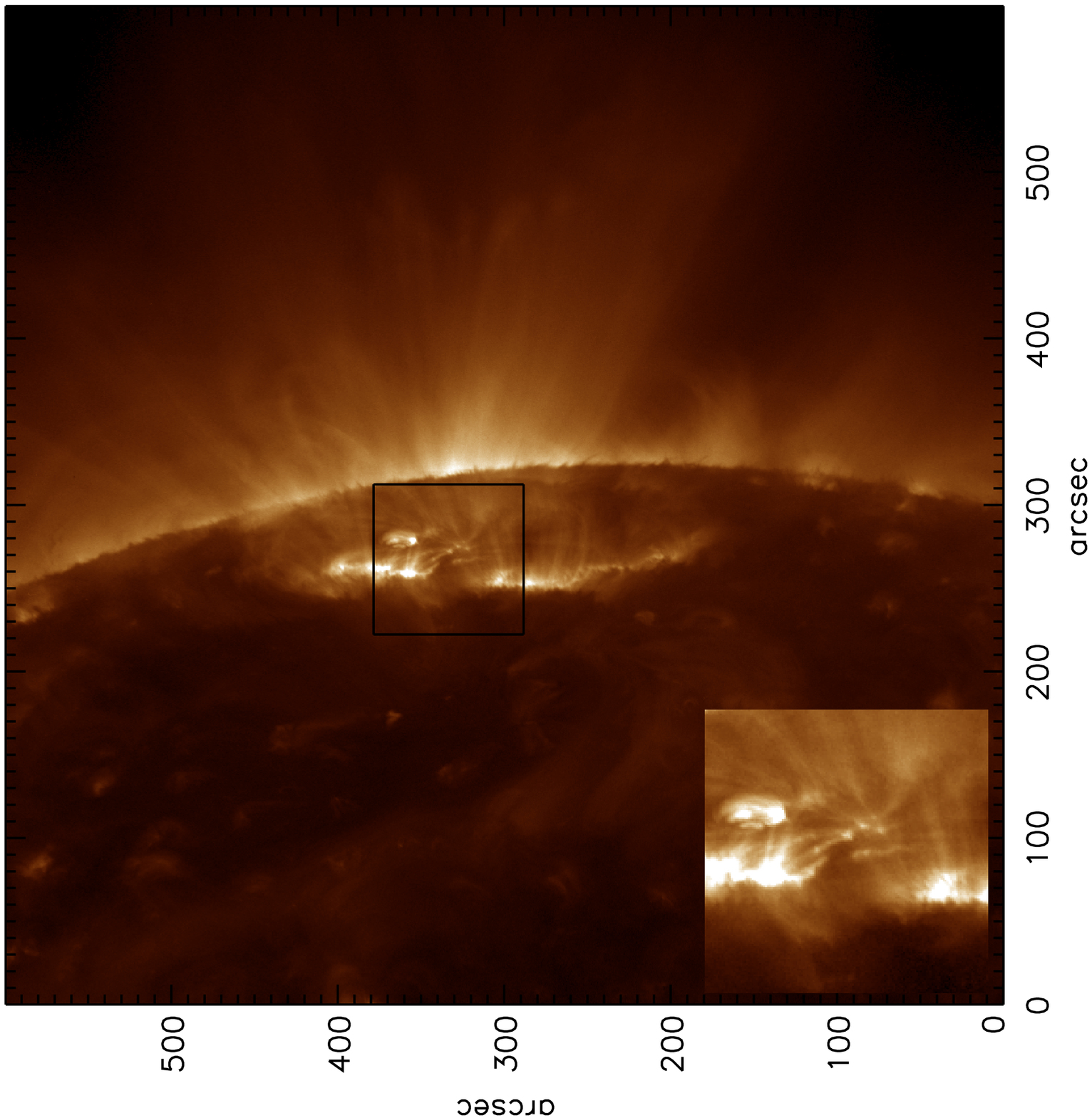}
}
\vspace{-5pt}
\caption{Left: Composite of contrast enhanced AIA 193\,\AA\, channel image and an HMI continuum intensity image 
with the white line tracing the loop starting from the LB. The inset on the lower right shows the magnified 
view of the loop extending down to the LB. Right: AIA 193\,\AA\,image for the same FOV as the left panel. The images
have been rotated for a better viewing perspective.}
\label{app_fig03}
\end{figure*}

Figure~\ref{app_fig03} shows the presence of coronal loops over AR 12741 when the spot was very close 
to the western limb. Using an unsharp image from the AIA 193\,\AA\, channel, we manually detect loops 
connecting the spot to the following polarity that appear as the extended network flux to the east and southeast
of the AR. The trace along one such set of loops along with the height is shown in the left panel of the figure.
   
\end{document}